\documentclass[journal]{IEEEtran}

\ifCLASSINFOpdf
\else
\fi

\usepackage{graphicx}
\usepackage{bm}
\usepackage{amsmath}
\usepackage{flushend}
\usepackage{cite}
\DeclareMathOperator{\sinc}{sinc}

\begin{document}

\title{Free space transmission lines in receiving antenna operation}

\author{
    \IEEEauthorblockN{Reuven Ianconescu\IEEEauthorrefmark{1},Vladimir Vulfin\IEEEauthorrefmark{2}}\\

    \IEEEauthorblockA{\IEEEauthorrefmark{2}Shenkar College of Engineering and Design, Ramat Gan, Israel, riancon@gmail.com}\\
    \IEEEauthorblockA{\IEEEauthorrefmark{1}Ben-Gurion University of the Negev, Beer Sheva 84105, Israel, vlad2042@yahoo.com}
}

\maketitle

\begin{abstract}
This work derives exact expressions for the voltage and current induced into a two conductors non isolated
transmission lines by an incident plane wave. The methodology is to use the transmission line radiating
properties to derive scattering matrices and make use of reciprocity to derive the response to the incident wave.
The analysis is in the frequency domain and it considers transmission lines of any small electric cross
section, incident by a plane wave from any incident direction and any polarisation.
The analytic results are validated by successful comparison with ANSYS commercial software simulation
results, and compatible with other published results.
\end{abstract}


%
\IEEEpeerreviewmaketitle

\section{Introduction}

\begin{figure}[!tbh]
\includegraphics[width=9cm]{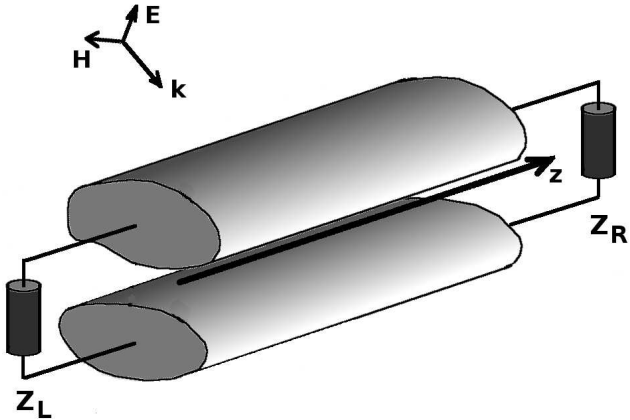}
\caption{Configuration of a two ideal conductors transmission line (TL), connected at both sides
to passive loads: $Z_L$ (left) and $Z_R$ (right), hit by a monochromatic plane wave propagating
toward the centre of coordinates.
The cross section is electrically small and may be of any shape.
The loads are located at the terminations of the TL, and are shown farther away, because of technical drawing
limitations.}
\label{config}
\end{figure}

This work calculates the voltage and differential current developed on an ideal
two-conductors TEM transmission line (TL) of any small electric cross section, connected
to passive (lumped) loads and hit by a monochromatic plane wave, as shown in Figure~\ref{config}.
We derive both amplitude and phase for the voltage and current developed along the TL, and hence
the powers delivered to the passive loads.

The configuration as defined here is a scattering problem, requiring a full wave solution to set
the tangential component of the electric field to 0 on the surface of the TL. However, we shall formulate
here an analytic solution to this problem, which is compared with a full wave HFSS solution.

\begin{figure}[!tbh]
\includegraphics[width=9cm]{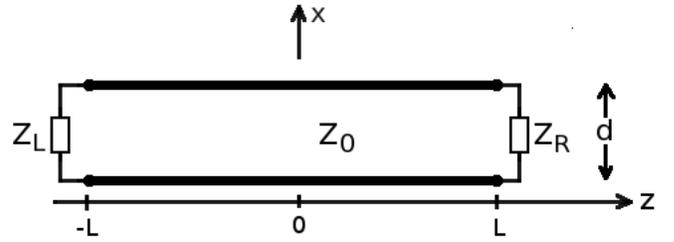}
\caption{Twin lead equivalent of the analysed transmission line, defined by the separation distance $d$ between the
conductors, and the characteristic impedance $Z_0$, both computed by the cross section analysis described in Appendix~B
of \cite{full_model_arxiv}. The TL is of length $2L$. }
\label{twin_lead_equivalent}
\end{figure}

As shown in \cite{full_model_arxiv}, the characteristic impedance $Z_0$ of the TL  and
an equivalent separation distance $d$ can be found by an electrostatic cross section analysis,
leading to a twin lead equivalent (examples for determining $Z_0$ and $d$ are shown in Appendix~B
of \cite{full_model_arxiv}). The twin lead equivalent is shown in Figure~\ref{twin_lead_equivalent}.
When the TL is excited at its port(s), this equivalent twin lead radiates the same far fields as the actual
TL. To emphasise, we {\it do not} solve for the twin lead geometry, but for an arbitrary cross section in
Figure~\ref{config}, the twin lead is only used as an intermediate tool for the calculations. 

The incident plane wave shown in Figure~\ref{config} propagates toward the coordinates origin with phase
$e^{-j\mathbf{k}\cdot\mathbf{r}}$, so that the wavenumber vector $\mathbf{k}=-\mathbf{\widehat{r}}k$ points toward
the origin. Expanding $\mathbf{\widehat{r}}$
in Cartesian unit vectors, the phase can be written as
\begin{equation}
e^{jk[x\sin\theta\cos\varphi+y\sin\theta\sin\varphi+z\cos\theta]},
\label{plane_wave_phase}
\end{equation}
where $\theta$ and $\varphi$ are the spherical angles which represent the direction of the plane wave arrival.

\begin{figure}[!tbh]
\includegraphics[width=9cm]{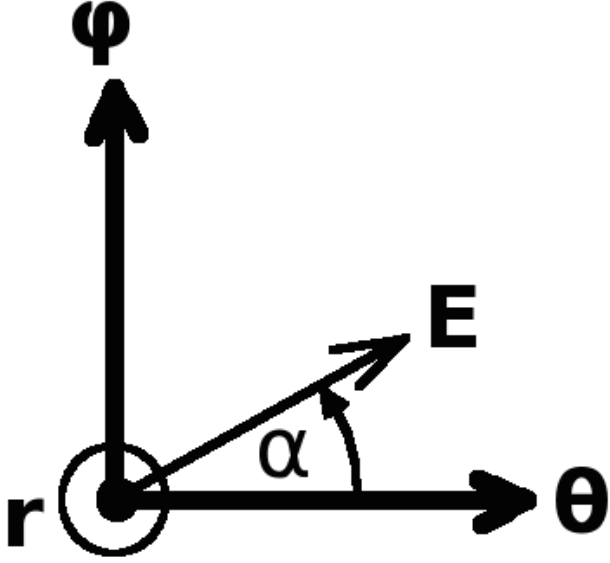}
\caption{The incident plane wave propagates toward the centre of coordinates in the $-\mathbf{\widehat{r}}$
direction. In 
spherical coordinates the local equiphase surface is $\theta,\varphi$ and the
polarisation is at angle $\alpha$ from the $\theta$ axis, so that at the origin
$\mathbf{E}=E_0(\boldsymbol{\widehat{\theta}}\cos\alpha+\boldsymbol{\widehat{\varphi}}\sin\alpha)$ (see
Eq.~(\ref{E_plane_wave})).}
\label{plane_wave}
\end{figure}

The polarisation of the incident plane wave is shown in Figure~\ref{plane_wave}.
It travels toward the centre of coordinates, perpendicular to the $\theta,\varphi$ plane,
with a polarisation angle $\alpha$ from the $\theta$ axis, so that the E field at the origin, where its phase
is 0 (see Eq.~(\ref{plane_wave_phase})) is given by
\begin{equation}
\mathbf{E}=E_0(\boldsymbol{\widehat{\theta}}\cos\alpha+\boldsymbol{\widehat{\varphi}}\sin\alpha)
\label{E_plane_wave}
\end{equation}
or its components
\begin{equation}
E_{\theta}=E_0\cos\alpha \,\,\,\,\,\, ; \,\,\,\,\,\, E_{\varphi}=E_0\sin\alpha.
\label{E_plane_wave_components}
\end{equation}

Such problem has been handled in \cite{Taylor,Smith,Harrison,Paul,Agrawal}, but not for an arbitrary
cross section (as in Figure~\ref{config}), and not for a general incident plane wave
(as in Figure~\ref{plane_wave}). The methodology used in these works is based on a generalisation
of the telegraph equations for transmission lines, for the twin lead geometry indicated in
Figure~\ref{twin_lead_equivalent}. Given $C'$ and $L'$ the capacitance and inductance per unit length,
using $Z_0=\sqrt{L'/C'}$ and the identity $\sqrt{L'C'}=1/c$, we write the (inhomogeneous) telegraph equations in the
following compact form
\begin{equation}
\frac{dV}{dz}+jk (Z_0I)=V_s
\label{telegraph1}
\end{equation}
\begin{equation}
Z_0\frac{dI}{dz}+jk V=Z_0I_s
\label{telegraph2}
\end{equation}
where the source terms $V_s$ and $I_s$ are induced by $H_y$ and $E_x$ fields of the incident plane wave,
integrated along the $x$ axis of the twin lead (see Figure~\ref{twin_lead_equivalent}), as follows:
\begin{equation}
V_s=jk\eta_0 \int_0^d H_y(x,z)dx\simeq jk\eta_0 H_y(z)d
\label{Vs}
\end{equation}
\begin{equation}
Z_0I_s=-jk\int_0^d E_x(x,z)dx\simeq -jk E_x(z)d,
\label{Is}
\end{equation}
and $\eta_0=377\Omega$ is the free space impedance. The last ($\simeq$) parts of Eqs.~(\ref{Vs}) and (\ref{Is})
are consistent with our framework which is restricted to small electric cross section ($kd\ll 1$).

The coupled non-homogeneous differential equations (\ref{telegraph1}) and (\ref{telegraph2}) can be decoupled into
two second order equations:
\begin{equation}
\frac{d^2V}{dz^2}+k^2 V=\frac{dV_s}{dz}-jk(Z_0I_s)
\label{sec_order_V}
\end{equation}
\begin{equation}
Z_0\frac{d^2I}{dz^2}+k^2 Z_0I=Z_0\frac{dI_s}{dz}-jkV_s
\label{sec_order_I}
\end{equation}
In \cite{Taylor} the authors wrote the general solution for those second order differential equations,
subject to the termination conditions:
\begin{equation}
V(-L)=-Z_LI(-L)\,\,\,\,\,\, ; \,\,\,\,\,\,V(L)=Z_RI(L),
\label{term_conditions}
\end{equation}
however they used in the expression for the sources the implicit assumption that the incident field
has only a $H_y$ component. In \cite{Smith} this solution has been simplified, and in \cite{Harrison,Paul,Agrawal} this
formalism has been extended to multiconductor TL.

\begin{figure}[!tbh]
\includegraphics[width=9cm]{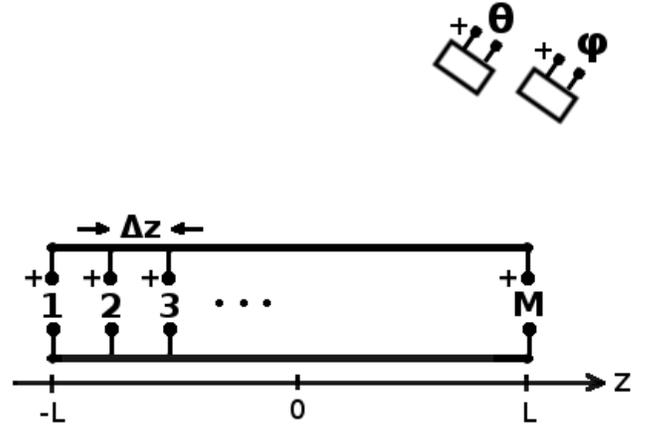}
\caption{$M$ parallel ports along the TL, adjacent ports are at distance $\Delta z$. Ports 1 and $M$ are
defined for the TL impedance $Z_0$, and the middle ports 2 .. $M-1$ are defined for a high impedance $Z_H$.
Two additional ports representing far antennas matched for the $\boldsymbol{\widehat{\theta}}$ and
$\boldsymbol{\widehat{\varphi}}$ polarizations are defined. Those ports are named $\theta$ and $\varphi$,
and are defined for the TL impedance $Z_0$.}
\label{voltage_S}
\end{figure}

\begin{figure}[!tbh]
\includegraphics[width=9cm]{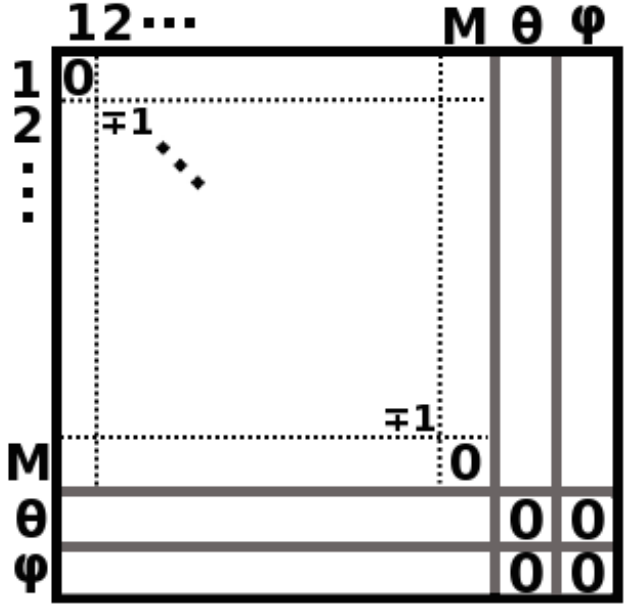}
\caption{Schematic diagram of a $(M+2)\times (M+2)$ scattering matrix describing the systems defined in
Figures~\ref{voltage_S} and \ref{current_S}. The first ports 1.. M are the TL ports in
the figures, and the two additional ports are the far $\theta$ or $\varphi$ polarised
antennas, named $\theta$, $\varphi$. In both cases
$S_{\theta\varphi}=S_{\varphi\theta}=0$, because they are orthogonally polarised, and
$S_{11}=S_{MM}=S_{\theta\theta}=S_{\varphi\varphi}=0$. As will be shown, the elements $S_{ii}$ for
$1<i<M$ are $\mp 1$ for the matrices defined by the configurations in Figures~\ref{voltage_S} and \ref{current_S},
respectively.}
\label{matrix}
\end{figure}

The results derived in this work satisfy Eqs.~(\ref{telegraph1})-(\ref{term_conditions}), but we use a different
technique to derive them. We shall use our
knowledge on the radiation properties of TL \cite{full_model_arxiv} to derive the
receiving properties, i.e. the response of a TL to an incident monochromatic plane wave described in
Figure~\ref{plane_wave}. To determine the voltage along the TL, we segment it into $M$ parallel ports
as shown in Figure~\ref{voltage_S}.
The ports 1 and $M$ are defined for the TL impedance $Z_0$, and the middle ports 2 .. $M-1$ are defined for a
high impedance $Z_H\to\infty$. We define two additional ports representing far antennas matched for the
$\boldsymbol{\widehat{\theta}}$ and $\boldsymbol{\widehat{\varphi}}$ polarizations of the field radiated by
the TL, defining a system of $M+2$ ports. The additional ports, $M+1$ and $M+2$, named $\theta$ and $\varphi$,
may be defined for any impedance, and for convenience we choose to define them for the TL impedance $Z_0$.
The scattering matrix of this system is shown schematically in Figure~\ref{matrix}.
Using the formalism developed in \cite{full_model_arxiv}, we know the $\boldsymbol{\widehat{\theta}}$ and
$\boldsymbol{\widehat{\varphi}}$ components of the radiated electric field (summarised in Appendix~\ref{normalization}).
Those field components are translated into voltages
using a normalisation explained in Appendix~\ref{normalization}, enabling us to calculate the generalised S matrix
(see Appendix~\ref{Generalised_S_matrix}) describing this system.

Here we excite this system with an incident voltage, scaled to the incident electric field in
Eq.~(\ref{E_plane_wave}) by
\begin{equation}
V^+=E_0\,d,
\label{Voltage_E_field_equivalence}
\end{equation}
which defines the voltage components (equivalent to Eq.~(\ref{E_plane_wave_components}))
\begin{equation}
V^+_{\theta}=V^+\cos\alpha \,\,\,\,\,\, ; \,\,\,\,\,\, V^+_{\varphi}=V^+\sin\alpha.
\label{Voltage_wave_components}
\end{equation}
We note that ports 1...~$M$ are matched with the $Z_0$ impedance at ports 1 and $M$ and $Z_H$ (i.e. infinite) on the
middle ports, which is exactly the configuration of a TL matched at both terminations, on which we want to obtain the voltages.

Those are obtained by multiplying the S matrix by the column vector
\begin{equation}
\left(
\begin{array}{c}
0\\
.\\
.\\
\\
0\\
\hline
V_{\theta}^+ \\
V_{\varphi}^+
\end{array}
\right)
\label{col_vec}
\end{equation}
of adequate excitation elements according to Eq.~(\ref{Voltage_wave_components}), 
and given $\Delta z$ is arbitrarily small (or $M$ arbitrarily large) we obtain the voltage as function of $z$
along the matched TL for the given incident plane wave. The solution for matched TL is after that generalised
for any terminations in Figure~\ref{config}.

\begin{figure}[!tbh]
\includegraphics[width=9cm]{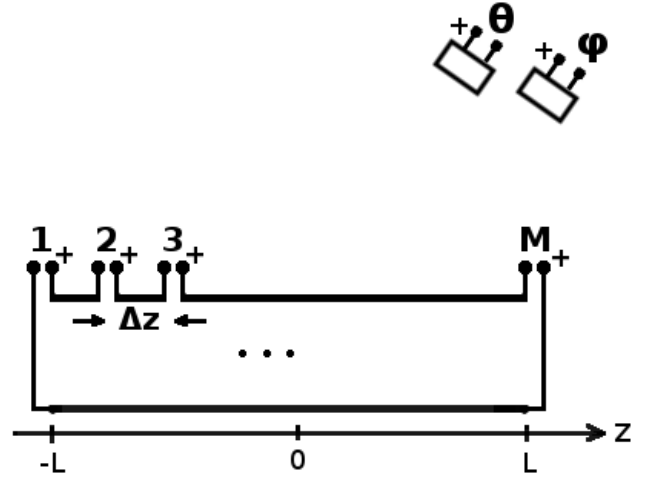}
\caption{$M$ serial ports along the TL, adjacent ports are at distance $\Delta z$. 
Ports 1 and $M$ are defined for the TL impedance $Z_0$, and the middle ports 2 .. $M-1$ are defined for a tiny
impedance $Z_T$. The ``+'' is defined at the right side of the port, so that the port current is defined in the
$+z$ direction. Note also that port 1 is physically identical to the one defined Figure~\ref{voltage_S} and so
is port $M$, up to the sign. Here we define as well the two additional ports named $\theta$ and $\varphi$,
defined for the TL impedance $Z_0$.}
\label{current_S}
\end{figure}

Similarly, the analytic solution for the current is obtained by segmenting the TL into $M$ serial ports
as shown in Figure~\ref{current_S}.
Ports 1 and $M$ are defined for the TL impedance $Z_0$, and the middle ports 2 .. $M-1$ are defined for a tiny
impedance $Z_T\to 0$. Here we define too the additional ports representing far antennas matched for the
$\boldsymbol{\widehat{\theta}}$ and $\boldsymbol{\widehat{\varphi}}$ polarizations of the field radiated by
the TL, obtaining a {\it different} system of $M+2$ ports.

For this system we also calculate a generalized S matrix, but defined for current waves (see Appendix~\ref{Generalised_S_matrix}).
Here the methodology is similar to the one explained for the voltages, the matched at both terminations TL result is
obtained using $Z_0$ impedance at ports 1 and $M$ and $Z_T$ (i.e. 0) on the middle ports, so we
first obtain the current on the matched TL and then generalise the result for any loads.

The advantages and novelties of our derivation are:
\begin{itemize}
\item The method of using the S matrix reciprocity to derive receiving characteristics from transmitting
characteristics is original and may be extended to additional configurations.
\item We do not limit ourselves to the twin lead cross section, presenting a general algorithm for TL of any
small electric cross section.
\item Our formulation is for a general incident plane wave, i.e. from an arbitrary direction $\theta$, $\varphi$, and
arbitrary polarisation $\alpha$.
\item The solution of Eqs.~(\ref{sec_order_V}) and (\ref{sec_order_I}) in \cite{Taylor} subject to general termination
conditions (\ref{term_conditions}) is extremely complicated, and it is not obvious in this method how one first
gets a simple solution for a matched at both terminations TL, and how to further generalise it. In our solution
based on S parameters, this emerges naturally.
\end{itemize}

The work is organized as follows. In Section~\ref{Voltage_S_matrix} we carry out all the analytic derivations
for the voltage and differential current along a TL hit by a monochromatic plane wave. We start with the voltage
on a TL with matched loads and generalise the result for a TL terminated in any loads ($Z_L$ and $Z_R$, as in
Figure~\ref{config}). This derivation is technical but quite lengthy,
therefore after deriving the voltage we give the final results for the current, which is calculated
in Appendix~\ref{Current_S_matrix}. Those calculations use the far E field radiated from
segments of the TL (based on \cite{full_model_arxiv} and summarised in Appendix~\ref{normalization}) and
the properties of the generalised scattering matrix summarised in Appendix~\ref{Generalised_S_matrix}.

In Section~\ref{HFSS} we describe the full wave HFSS simulations performed, and explain some delicate issues regarding
the measurements of voltage and current. To be mentioned that a TL hit by a monochromatic plane wave may develop common
mode currents. However, our derivation of the current is based on the reciprocity of the system defined
in Figure~\ref{current_S}. The currents used to define the scattering matrix are differential, and for those
differential currents we use the far fields developed in \cite{full_model_arxiv}. Hence, when we use the reciprocity
to express the current developed on the TL as function of the plane wave excitation, we obtain only the differential
part of the current, and this drawback applies also to the method used in \cite{Taylor,Smith,Harrison,Paul,Agrawal}.
If the purpose is to calculate the power on the loads, this is not a limitation, because
the common mode does not affect this power, but this has to be taken into account in the current measurement
in Section~\ref{HFSS}.

In Section~\ref{validation_hfss} we validate the analytic results with full wave
HFSS simulations for a cross section which is not a twin lead, showing that this formalism works for a general electrically
small cross section. We compare the theoretical results obtained in Sections~\ref{Voltage_S_matrix}
for the voltage and differential current with full wave HFSS solutions for matched
and non matched transmission lines. In Section~\ref{validation_compatibility} we prove the analytical results obtained
in Section~\ref{Voltage_S_matrix} are fully compatible with the results obtained in \cite{Taylor,Smith,Harrison,Paul,Agrawal}
and therefore satisfy Eqs.~(\ref{telegraph1})-(\ref{term_conditions}).

In Section~\ref{power} we calculate general expressions for the powers transferred to the loads, and interpret them
in terms of the transmission radiation patterns in \cite{full_model_arxiv} and incident plane wave polarisation. We
show that the power in the ``left'' load is closely related to the transmitting properties of a source at the left termination
and viceversa. In Section~\ref{transmit_receive} we show a full analysis of the connection between the radiation
characteristics of TL calculated in \cite{full_model_arxiv} and the receiving characteristics calculated in this
work.

The work is ended with some concluding remarks.

Note: through this work, the phasor amplitudes are RMS values, hence there is no 1/2 in the expressions for power.
Also, it is worthwhile to mention that the results of this work depend on physical sizes relative to the wavelength,
and hence are valid for all frequencies satisfying the condition of small electric cross section.

\section{Derivation of the voltage and differential current along the TL}
\label{Voltage_S_matrix}

We start with the voltages, analysing the system described in Figure~\ref{voltage_S}.

As explained in the introduction, to find the voltages on a matched TL, i.e. 
for $Z_L=Z_R=Z_0$ in Figure~\ref{twin_lead_equivalent}, one needs only the
cross elements between the group of ports 1...~$M$ and the group $\theta$, $\varphi$ of the S matrix.

However, to generalise the solution for the voltage along a TL loaded by and $Z_L$ and $Z_R$, one
needs additional elements of the S matrix, we therefore calculate here all the elements of the
matrix for the system defined in Figure~\ref{voltage_S}, schematically shown in Figure~\ref{matrix}.
This matrix satisfies
\begin{equation}
\mathbf{V}^-=\mathbf{S} \mathbf{V}^+,
\label{S_for_voltages}
\end{equation}
where $\mathbf{S}$ is an $(M+2)\times(M+2)$ matrix and $\mathbf{V}^+$, $\mathbf{V}^-$ are
column vectors of incoming and outgoing voltages, respectively.

We start with the submatrix with indices 1 to $M$. Feeding a {\it middle} port $1<n<M$, located at
\begin{equation}
z_n=-L+(n-1)\Delta z,
\label{z_k}
\end{equation}
where
\begin{equation}
\Delta z=\frac{2L}{M-1}
\label{delta_z}
\end{equation}
with a forward (entering) voltage $V^+_n$ and terminating the other
TL ports $1\le i\le M$ by the impedances defined for those ports (i.e. $Z_0$ for ports 1 and $M$, and $Z_H$ for the middle ports)
results in a forward wave from $z=z_n$ to $z=L$ and a backward wave from $z=z_n$ to $z=-L$, because the waves
encounter at the intermediate ports a very high impedance $Z_H\to\infty$.

The impedance seen at port $n$ is $Z_0/2$, so the reflection coefficient is $S_{n,n}=\frac{Z_0/2-Z_H}{Z_0/2+Z_H}\simeq -1+\frac{Z_0}{Z_H}$,
therefore the port voltage is $V_n=V^+_n(1+S_{n,n})=V^+_nZ_0/Z_H$. The outgoing voltages at ports $i\neq n$ are
\begin{equation}
V^-_{i\neq n}=V_ne^{-jk\Delta z|i-n|}=V^+_n\frac{Z_0}{Z_H}e^{-jk\Delta z|i-n|}.
\label{V_minus_i_neq_k}
\end{equation}
and for $i=n$
\begin{equation}
V^-_n=V^+_nS_{n,n}=V^+_n\left(-1+\frac{Z_0}{Z_H}\right)
\label{V_minus_kk}
\end{equation}

This results in the following (partial) $n$ column of the S matrix
\begin{equation}
S_{1\le i\le M\,,\,1<n<M}=
\left\{
\begin{array}{l l}
\frac{Z_0}{Z_H}e^{-jk\Delta z|i-n|} &  \,\, i\neq n  \\
-1+\frac{Z_0}{Z_H}\simeq -1     &  \,\, i=n
\end{array}
\right.,
\label{partial_column_k}
\end{equation}
see upper sign in Figure~\ref{matrix}. For column $n=1$ or $M$, the results are similar,
only replace $\frac{Z_0}{Z_H}$ by 1, hence
\begin{equation}
S_{1\le i\le M\,,\,n=1,M}=
\left\{
\begin{array}{l l}
e^{-jk\Delta z|i-n|} &  \,\, i\neq n  \\
0                &  \,\, i=n
\end{array}
\right.
\label{partial_column_k_eq_1_or_M}
\end{equation}

Eqs.~(\ref{partial_column_k}) and (\ref{partial_column_k_eq_1_or_M}) define the upper left square of
the S matrix, i.e. all the elements connecting the TL ports 1 to $M$ - see Figure~\ref{matrix}, on which
the diagonal elements are shown, the upper sign relates to this system.

Now we add the columns $S_{\theta,n}$ and $S_{\varphi,n}$ (last columns in Figure~\ref{matrix}),
representing the far $\boldsymbol{\widehat{\theta}}$ and $\boldsymbol{\widehat{\varphi}}$
polarisation antennas, respectively.

Feeding a {\it middle} port $1<n<M$ with the voltage $V^+_n$ and all other ports by their matched load, we calculate the far fields
given in Eqs.~(\ref{E_theta_x})-(\ref{E_theta_plus_minus_z}), scaled to the voltages $V^-_{\theta}$ and $V^-_{\varphi}$
according to Eq.~(\ref{scale_factor}).

\begin{figure}[!tbh]
\includegraphics[width=9cm]{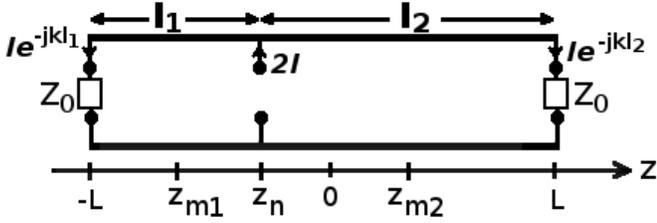}
\caption{Currents in the TL while feeding port $n$ with $V^+_n$. Defining $I\equiv V^+_n/Z_H$, the current in the feeding
line at $z=z_n$ is $2I$ in the positive $x$ direction, and the currents at $z=L$ (port $M$) and $z=-L$ (port 1) are
$Ie^{-jkl_2}$ and $Ie^{-jkl_1}$,
respectively, both in the negative $x$ direction. The $z$ directed current waves are the forward wave in the region $[z_n,L]$
and the backward current wave in the region $[-L,z_n]$, to be used in
Eq.~(\ref{E_theta_plus_minus_z}). For the forward wave we use the current at the middle point
$z_{m2}=(z_n+L)/2$, which is $Ie^{-jkl_2/2}$, while for the backward wave we use the current at $z_{m1}=(z_n-L)/2$:
$-Ie^{-jkl_1/2}$ (minus because it flows in the $-z$ direction).}
\label{currents_configuration}
\end{figure}

We start with the contribution of the $x$ currents given in Eqs.~(\ref{E_theta_x}) and (\ref{E_phi_x}), 
see Figure~\ref{currents_configuration}.
It is convenient to express the results using the distances from the feeding point to the terminations:
\begin{equation}
l_2=L-z_n\,\,\,\,\,\, ; \,\,\,\,\,\,l_1=z_n+L,
\label{l_12_k}
\end{equation}
shown in Figure~\ref{currents_configuration}, where $z_n$ is defined in Eq.~(\ref{z_k}).
Hence in all following expressions, the value of $n$ is ``hidden'' in $l_1$ and $l_2$. To shorten the
notations in the forthcoming results we introduce two definitions
\begin{align}
&f_1\equiv e^{-jk(l_1+l_2\cos\theta)/2}\sin(kl_1\cos^2(\theta/2)) \notag \\
&f_2\equiv e^{-jk(l_2-l_1\cos\theta)/2}\sin(kl_2\sin^2(\theta/2)).
\label{f_12}
\end{align}
Using Eqs.(\ref{E_theta_x}) for each of the three currents
at $z_w=z_n$, $L$ and $-L$, scaling with (\ref{scale_factor}), after some algebra we obtain:
\begin{equation}
V^-_{\theta \,(x)}=-jV^+_n(Z_0/Z_H)[f_1+f_2]\cos\theta\cos\varphi 
\label{V_minus_theta_x_1}
\end{equation}
As evident from Eqs.(\ref{E_theta_x}) and (\ref{E_phi_x}), $V^-_{\varphi \,(x)}$ is identical to the above, up to replacing $\cos\theta\cos\varphi$
by $-\sin\varphi$, we therefore have
\begin{equation}
V^-_{\varphi \,(x)}=jV^+_n(Z_0/Z_H)[f1+f_2]\sin\varphi.
\label{V_minus_phi_x}
\end{equation}
Next we calculate the contributions of the forward current wave in the region $z=[z_n,L]$ and the backward current wave
in the region $z=[-L,z_n]$ (see Figure~\ref{currents_configuration}), using the upper and lower signs in
Eq.~(\ref{E_theta_plus_minus_z}), respectively. For the forward wave we use the current in the middle point
$z_{m2}=(z_n+L)/2$, having the phase $e^{-jkl_2/2}$ for a TL length $h=l_2$, while for the backward wave we use the current at $z_{m1}=(z_n-L)/2$,
having the phase $e^{-jkl_1/2}$, for a TL length $h=l_1$. Also the backward current is defined in the $-z$ direction, it therefore is taken with the
minus sign. The middle points can be written more compactly as
\begin{equation}
z_{m2}=l_1/2\,\,\,\,\,\, ; \,\,\,\,\,\,z_{m1}=-l_2/2,
\label{z_m12}
\end{equation}
and after scaling with (\ref{scale_factor}), we obtain:
\begin{equation}
V^-_{\theta \,(z)}=2jV^+_n(Z_0/Z_H)\cos\varphi[f_2\cos^2(\theta/2) - f_1\sin^2(\theta/2) ]
\label{V_minus_theta_z}
\end{equation}
Now we add the $x$ and $z$ directed currents contributions to $V^-_{\theta}=V^-_{\theta \,(x)}+V^-_{\theta \,(z)}$, using $1-\cos(\theta)=2\sin^2(\theta/2)$
and $1+\cos(\theta)=2\cos^2(\theta/2)$, after some algebra we obtain:
\begin{equation}
V^-_{\theta}=jV^+_n(Z_0/Z_H)[f_2-f_1]\cos\varphi
\label{V_minus_theta}
\end{equation}
$V^-_{\varphi}$ has only the contribution of the $x$ directed currents, given in Eq~(\ref{V_minus_phi_x}), hence
\begin{equation}
V^-_{\varphi}=V^-_{\varphi \,(x)}.
\label{V_minus_phi}
\end{equation}
Results (\ref{V_minus_theta}) and (\ref{V_minus_phi}) define the $S_{\theta,n}$ and
$S_{\varphi,n}$ matrix elements respectively, for the columns $1<n<M$:
\begin{equation}
S_{\theta,\,1<n<M}=j(Z_0/Z_H)[f_2-f1]\cos\varphi
\label{S_theta_k}
\end{equation}
\begin{equation}
S_{\varphi,\,1<n<M}=j(Z_0/Z_H)[f_1+f_2]\sin\varphi.
\label{S_phi_k}
\end{equation}
For column $n=1$ or $M$, the results are similar, only replace $\frac{Z_0}{Z_H}$ by 1, hence
\begin{equation}
S_{\theta,\,n=1,M}=j[f_2-f_1]\cos\varphi
\label{S_theta_k_1_M}
\end{equation}
\begin{equation}
S_{\varphi,\,n=1,M}=j[f_1+f_2]\sin\varphi.
\label{S_phi_k_1_M}
\end{equation}
The transpose elements are found by the reciprocity condition $S_{i,j}Z_j=S_{j,i}Z_i$ (see
Appendix~\ref{Generalised_S_matrix}, Eq.~\ref{GSM_reciprocity}), where $Z_i$ and
$Z_j$ are the impedances for which ports $i$ and $j$ have been defined, respectively. Given the ports
1,$M$,$\theta$ and $\varphi$ are defined for $Z_0$ of the TL and ports 2 .. $M-1$ are defined for
$Z_H$, the transposed relations $S_{n,\theta}$ and $S_{n,\varphi}$ are given by Eqs.~(\ref{S_theta_k_1_M})
and (\ref{S_phi_k_1_M}), for all $n$:
\begin{equation}
S_{n,\theta}=j[f_2-f_1]\cos\varphi
\label{S_k_theta}
\end{equation}
\begin{equation}
S_{n,\varphi}=j[f_1+f_2]\sin\varphi.
\label{S_k_phi}
\end{equation}
When issuing and excitation $V^+_{\theta}$ from the $\theta$ port and matching all other ports, the voltages
at ports $n$ on the TL are $V^-_n=V^+_{\theta}S_{n,\theta}$. Given the number $M$ is not limited, one can get
the continuous voltage on the TL for the excitation of the $\theta$ port:
\begin{equation}
V_{\theta}(z)=jV^+_{\theta}[f_2(z)-f_1(z)]\cos\varphi
\label{V_theta}
\end{equation}
where the $z$ dependence in $f_{1,2}$ (see Eq.~(\ref{f_12})) is in $l_1$ and $l_2$, defined according to
Eq.~(\ref{l_12_k}), using $z$ for $z_n$.
This voltage develops on the TL due to a $\boldsymbol{\widehat{\theta}}$ polarised plane wave coming from
coordinates $(\theta,\varphi)$ at infinity.

Similarly for an excitation $V^+_{\varphi}$ from the $\varphi$ port, while matching all other ports, one obtains
the continuous voltage on the TL:
\begin{equation}
V_{\varphi}(z)=jV^+_{\varphi}[f_1(z)+f_2(z)]\sin\varphi.
\label{V_phi}
\end{equation}
which develops due to a $\boldsymbol{\widehat{\varphi}}$ polarised plane wave coming from
coordinates $(\theta,\varphi)$ at infinity.

The total voltage developed on the matched TL is $V_{\theta}(z)+V_{\varphi}(z)$, where the intensities of
$V^+_{\theta}$ and $V^+_{\varphi}$ are according to Eq.~(\ref{Voltage_wave_components}). It comes out
\begin{equation}
V(z)=jV^+[f_2\cos(\varphi-\alpha)-f_1\cos(\varphi+\alpha)]
\label{V_tot}
\end{equation}

Eq.~(\ref{V_tot}) is the final result for the voltage on a matched TL, i.e. terminated at both ends with
resistors $Z_0$. We now generalize the above result for a TL terminated with any loads at ports 1 and $M$:
$Z_L$ (left) and $Z_R$ (right), respectively (see Figure~\ref{twin_lead_equivalent}).
We express the generalized results in terms of the reflection coefficients:
\begin{equation}
\Gamma_L=\frac{Z_L-Z_0}{Z_L+Z_0}\,\,\,\text{and}\,\,\,\Gamma_R=\frac{Z_R-Z_0}{Z_R+Z_0}.
\label{reflection_coeffs}
\end{equation}
For the matched case, the only incoming voltage waves are $V^+_{\theta}$ and $V^+_{\varphi}$, while all other
$V^+_{1\le i\le M}=0$. Here we have two additional incoming voltages at the mismatched ports 1 and $M$:
\begin{equation}
V^+_1=\Gamma_L V^-_1\,\,\,\text{and}\,\,\,V^+_M=\Gamma_R V^-_M,
\label{V_plus_1_M}
\end{equation}
while at the matched (with $Z_H$) ports $n=2..\,M-1$ the incoming voltages are 0:
\begin{equation}
V^+_{2\le n\le M-1}=0.
\label{V_minus_2_M_minus_1}
\end{equation}
We use the general connection for $1\le n\le M$ (i.e. on TL) but $1\le i\le M+2$,
i.e. all excitations including $\theta$ and $\varphi$:
\begin{align}
  V^-_n=&\sum_{i=1}^{M+2} S_{n,i}V^+_i=S_{n,1}V^+_1+S_{n,M}V^+_M+V(z)= \notag \\
       &S_{n,1}\Gamma_L V^-_1+S_{n,M}\Gamma_R V^-_M+V(z)
\label{V_minus_k}
\end{align}
where the only non zero terms are $i=1,M,\theta$ and $\varphi$ (see Eq~(\ref{V_minus_2_M_minus_1})),
and the last two terms $i=\theta,\varphi$ represent the matched voltage at $z=z_n$, given in (\ref{V_theta}),
(\ref{V_phi}), summed in (\ref{V_tot}). In the last form of Eq.~(\ref{V_minus_k}), we used (\ref{V_plus_1_M}).

For $n=1$, Eq.~(\ref{V_minus_k}) becomes
\begin{equation}
V^-_1=\Gamma_R V^-_M e^{-jk2L} + V(-L).
\label{V_minus_1_1}
\end{equation}
because $S_{1,1}=0$, $S_{1,M}=S_{M,1}=e^{-jk2L}$ and $z_n=-L$.

For $n=M$, Eq.~(\ref{V_minus_k}) reads
\begin{equation}
V^-_M=\Gamma_L V^-_1 e^{-jk2L} + V(L).
\label{V_minus_M_1}
\end{equation}
because $S_{M,M}=0$ and $z_n=L$.

Using $S_{n,1}=e^{-jkl_1}$, $S_{n,M}=e^{-jkl_2}$, for $2\le n\le M-1$, Eq.~(\ref{V_minus_k}) becomes
\begin{equation}
V^-_{2\le n\le M-1}= e^{-jkl_1}\Gamma_L V^-_1 + e^{-jkl_2} \Gamma_R V^-_M + V(z),
\label{V_minus_k_between_1_and_M_1}
\end{equation}
where in principle $V(z)$ here excludes the terminations, but as we shall see this exclusion is not necessary.
We solve now Eqs.~(\ref{V_minus_1_1}) and (\ref{V_minus_M_1}) for $V^-_1$ and $V^-_M$ and obtain:
\begin{equation}
V^-_1=\frac{\Gamma_R e^{-j2kL}V(L)+V(-L)}{1-\Gamma_L\Gamma_R e^{-j4kL}} 
\label{V_minus_1_solution}
\end{equation}
\begin{equation}
V^-_M=\frac{\Gamma_L e^{-j2kL}V(-L)+V(L)}{1-\Gamma_L\Gamma_R e^{-j4kL}}.
\label{V_minus_M_solution}
\end{equation}
To avoid confusions the generalised (non-matched) results are subindexed ``NM''.
The total (non matched) voltage $V_{\text{NM}}(z_n)=V^+_n+V^-_n$. For ports 1 or M, this voltage is
\begin{equation}
V_{\text{NM}}(-L)=V^-_1(1+\Gamma_L)
\label{V_NM_port_1}
\end{equation}
\begin{equation}
V_{\text{NM}}(L)=V^-_M(1+\Gamma_R).
\label{V_NM_port_M}
\end{equation}
For the ports $2\le n\le M-1$, $V^+_n=0$ (\ref{V_minus_2_M_minus_1}),
therefore Eq.~(\ref{V_minus_k_between_1_and_M_1}) describes the total voltage on those ports. Taking the limit
$z\to -L$ of Eq.~(\ref{V_minus_k_between_1_and_M_1}), using (\ref{V_minus_1_1}) we find it reduces to
(\ref{V_NM_port_1}) and similarly the limit $z\to L$ of Eq.~(\ref{V_minus_k_between_1_and_M_1}) reduces to
(\ref{V_NM_port_M}), therefore Eq.~(\ref{V_minus_k_between_1_and_M_1}) describes
the non matched voltage on the TL at all ports (in the continuum, for all $z$)
\begin{equation}
V_{\text{NM}}(z)=  V(z) + \Delta V(z),
\label{V_NM}
\end{equation}
where $\Delta V(z)$ is the correction term due to non matching (first two terms in (\ref{V_minus_k_between_1_and_M_1})):
\begin{equation}
\Delta V(z)= e^{-jkl_1}\Gamma_L V^-_1 + e^{-jkl_2} \Gamma_R V^-_M,
\label{Delta_V}
\end{equation}
and $V^-_1$ and $V^-_M$ are given in Eqs.~(\ref{V_minus_1_solution}) and (\ref{V_minus_M_solution}).

This concludes the voltage developed on the TL due to a monochromatic plane wave, for a TL matched at both ends,
the solution is Eq.~(\ref{V_tot}) and for general terminations we add the correction term in Eq.~(\ref{Delta_V}).

A similar calculation is carried out to find the differential current developed on the TL due to a monochromatic
plane wave. This is done by analysing the system defined in Figure~\ref{current_S}. Because this calculation is
lengthy, and similar to this carried out on the system defined in Figure~\ref{voltage_S}, it is done in
Appendix~\ref{Current_S_matrix}, and the results are given below.

The currents on a TL matched on both terminations due to $\theta$ or $\varphi$ polarisations are:
\begin{equation}
I_{\theta}(z)=-j(V^+_{\theta}/Z_0)[f_1(z)+f_2(z)]\cos\varphi
\label{I_theta_1}
\end{equation}
\begin{align}
I_{\varphi}(z)=j(V^+_{\varphi}/Z_0)[f_1(z)-f_2(z)]\sin\varphi,
\label{I_phi_1}
\end{align}
and the total current for matched TL is their sum:
\begin{equation}
I(z)=-j(V^+/Z_0)[f_1\cos(\varphi+\alpha)+f_2\cos(\varphi-\alpha)]
\label{I_tot}
\end{equation}
It is easy to check that Eqs.~(\ref{V_tot}) and (\ref{I_tot}) satisfy the termination conditions
(\ref{term_conditions}) with $Z_L=Z_R=Z_0$.

For a TL terminated in any impedances $Z_L$, $Z_R$, the differential current along the TL,
$I_{\text{NM}}$ (i.e. non matched) is
\begin{equation}
I_{\text{NM}}(z)= I(z) + \Delta I(z),
\label{I_NM}
\end{equation}
where $\Delta I(z)$ is the correction term due to non matching:
\begin{equation}
\Delta I(z)= -e^{-jkl_1}\Gamma_L I^-_1 - e^{-jkl_2} \Gamma_R I^-_M,
\label{Delta_I}
\end{equation}
and $I^-_1$ and $I^-_M$ are given explicitly in Eqs.~(\ref{I_minus_1_solution}) and (\ref{I_minus_M_solution}),
and also satisfy
\begin{equation}
V^-_1=-Z_0I^-_1\,\,\,\,\,\, ; \,\,\,\,\,\,V^-_M=Z_0I^-_M
\label{VI_minus_1_M}
\end{equation}

The solutions (\ref{V_NM}) and (\ref{I_NM}) satisfy the termination conditions
(\ref{term_conditions}).

\section{Full wave HFSS simulations}
\label{HFSS}

\begin{figure}[!tbh]
\includegraphics[width=8cm]{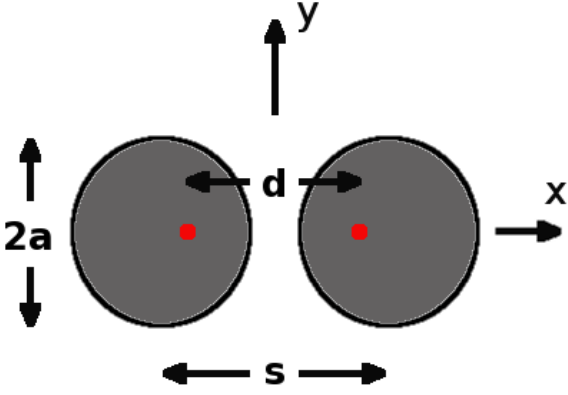}
\caption{Cross section of two parallel cylinders: the distance between the centres is $s=3.59$cm,
and the diameters are $2a=2.54$cm. The red points show the current images which 
define the twin lead representation, and the distance between them $d=2.537$cm is calculated in
Eq.~(\ref{d}).}
\label{parallel_cylinders}
\end{figure}

We describe in this Section the HFSS simulations done for the scattering problem defined in
Figure~\ref{config}. The results of this simulations are compared in the next section with the
analytic results. We used a non twin lead cross section (used also in \cite{full_model_arxiv}),
shown in Figure~\ref{parallel_cylinders}.
For this cross section one knows analytically the distance $d$ in the twin lead representation
(shown as red points in Figure~\ref{parallel_cylinders}) by image theory, yielding
\begin{equation}
d=\sqrt{s^2-(2a)^2}=2.537\text{cm},
\label{d}
\end{equation}
and also the characteristic impedance
\begin{equation}
Z_0=\frac{\eta_0}{\pi} \ln\left(\frac{d+s}{2a}\right)=105.6\,\Omega
\label{Z0}
\end{equation}
while for other cross sections those can be determined by an electrostatic ANSYS 2D “Maxwell”
simulation, as shown in Appendix~B of \cite{full_model_arxiv}.

The electric field of a plane wave is by default $E_0=1$V/m in the HFSS simulation, so to normalise the
results for $E_0\,d=1$V we divide the measured results by the value of $d$ in Eq.~(\ref{d}).

For convenience, we shall use a fixed TL length of $l\equiv 2L=125$cm, and test for different
frequencies. We measure the voltage and current along the TL from $z=-61.25$cm to $z=61.25$cm
at intervals of 6.125cm, in total at 21 points. At the TL terminations $z=-L$ and $z=L$, we
use inactive lumped ports defined for the impedance we need at those terminations.

Two-conductors transmission lines excited {\it only} at terminations, develop the TEM mode, so that both $E_z$ and $H_z$
are 0. In such case one can measure the voltage by $\int \mathbf{E}\cdot \mathbf{dl}$
from the ``+'' to the ``-'' conductor and the current using
$\oint \mathbf{H}\cdot \mathbf{dl}$ around the ``+'' conductor, both on {\it any} integration path.

In the case analyzed here, the TL is excited by an external plane wave, therefore, depending on the
incidence of this wave $E_z$ and/or $H_z$ are not necessarly 0, we therefore need more careful
definitions for the voltage and current measurements, as described in the following subsections.

\subsection{Voltage measurement}

\begin{figure}[!tbh]
\includegraphics[width=8cm]{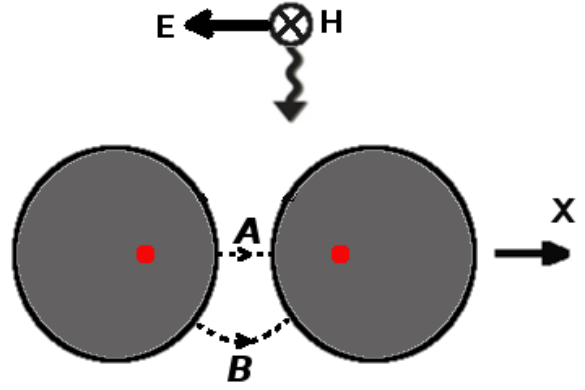}
\caption{Cross section voltage measurement on two possible pathes $A$ and $B$. In case $H_z=0$,
all pathes lead to the same result, namely $\int_A \mathbf{E}\cdot \mathbf{dl}=\int_B \mathbf{E}\cdot \mathbf{dl}$.
However, if $H_z\neq 0$, as for the incident plane wave shown here, the result of the path integral depends on
the paths used, and the correct voltage measurement is $\int_A \mathbf{E}\cdot \mathbf{dl}$, i.e. along the $x$ axis
consistent with the parallel ports definition in Figure~\ref{voltage_S}.}
\label{parallel_cylinders_Hz}
\end{figure}

The voltage measured by the integral $\int \mathbf{E}\cdot \mathbf{dl}$ in the cross
section depends on the chosen integration path if $H_z\neq 0$, as shown in Figure~\ref{parallel_cylinders_Hz}.
To define the correct path we look at the definitions of the parallel ports in Figure~\ref{voltage_S}.
Those have been defined on the $x-z$ plane, so that {\it only} $x$ directed currents flow through the port, and this fact
has been used in the calculation of the S matrix in Section~\ref{Voltage_S_matrix}, see
Figure~\ref{currents_configuration}.

Therefore, to be consistent with the parallel ports definition,
the correct path to measure the voltage is path $A$ (on the $x$ axis) shown in Figure~\ref{parallel_cylinders_Hz},
and this path is used in all the voltage measurements in Section~\ref{validation}.

\subsection{Current measurement}

\begin{figure}[!tbh]
\includegraphics[width=8cm]{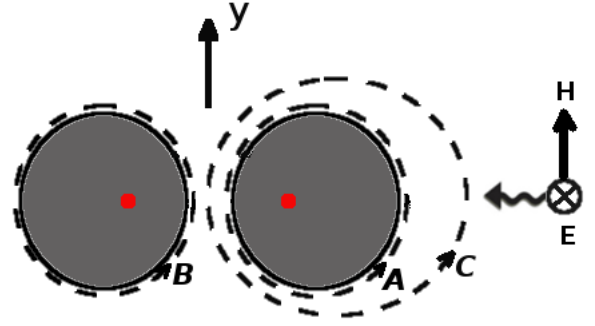}
\caption{Cross section differential current measurement for the case $H_z\neq 0$ requires integration on tight loops
around each conductor from which one obtains the currents $I_A$ and $I_B$ (Eq.~(\ref{IAB})).
The differential current is calculated in Eq.~(\ref{I_simul}).}
\label{parallel_cylinders_Ez}
\end{figure}

For the case $E_z=0$, the current can be measured by $\oint \mathbf{H}\cdot \mathbf{dl}$ around
the ``positive'' conductor, on any integration path, like path $C$ in Figure~\ref{parallel_cylinders_Ez}.
In the opposite case consisting in an
incident plane wave for which $E_z\neq 0$ (as shown in the figure), not only the integration path has to be tight around
the conductor, but also, due to common mode current, one has to measure around both conductors, on
paths $A$ and $B$, as shown in Figure~\ref{parallel_cylinders_Ez},
to obtain the currents
\begin{equation}
I_A=\oint_A \mathbf{H}\cdot \mathbf{dl}\,\,\,\,\,\, ; \,\,\,\,\,\,I_B=\oint_B \mathbf{H}\cdot \mathbf{dl}.
\label{IAB}
\end{equation}
As one notes from the derivation in Section~\ref{Current_S_matrix}, the S matrix has been derived from
differential currents, so the analytic results (\ref{I_theta_1}) and (\ref{I_phi_1}) represent differential currents.
To compare the HFSS simulation results to the analytic results we calculate the simulation differential
current:
\begin{equation}
I=(I_A-I_B)/2;
\label{I_simul}
\end{equation}
If $H_z=0$, $I_B$ trivially reduces to $-I_A$, so that the differential current is the current on the ``positive''
conductor, which may be calculated by integrating on any path around it, like paths $A$ or $C$ in
Figure~\ref{parallel_cylinders_Ez}.

\section{Validation of the analytic results}
\label{validation}
\subsection{Comparisons with HFSS full wave solution}
\label{validation_hfss}

\subsubsection{Matched transmission line}
\label{matched}

We validate in this section the analytic results for matched TL in Eqs~(\ref{V_tot})
and (\ref{I_tot}) by comparison with full wave solution of ANSYS HFSS simulation results,
described in the previous section.

We analyse three examples, each from a main incidence direction, by plane waves traveling along the $x$, $y$ or $z$
axes. In all examples the incident electric field intensity satisfies $Ed=\eta_0Hd=1$V. The half lenght of the TL
is $L=0.625$~m (or the length $2L=1.25$~m).

\begin{figure}[!tbh]
\includegraphics[width=9cm]{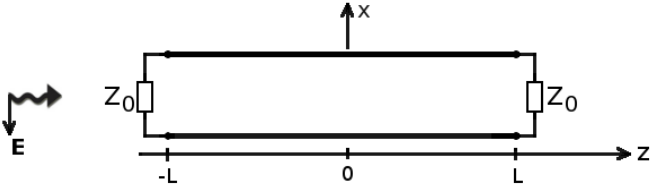}
\caption{Matched TL illuminated by a $-\mathbf{\widehat{x}}$ polarised plane wave from $\theta=\pi$.}
\label{left_incidence}
\end{figure}

In the first example we examine a plane wave traveling from $\theta=\pi$, along the $z$ axis,
colinear with the TL, having the phase $e^{-jkz}$, as shown in Figure~\ref{left_incidence}.
For $\theta=\pi$, Eqs.~(\ref{V_tot}) and (\ref{I_tot}) (normalized by $Z_0$) reduce to
\begin{equation}
V(z)=V^+je^{-jkL}\sin[k(L-z)]\cos(\varphi-\alpha)
\label{voltage_left_incidence_gen}
\end{equation}
\begin{equation}
Z_0I(z)=-V^+je^{-jkL}\sin[k(L-z)]\cos(\varphi-\alpha),
\label{current_left_incidence_gen}
\end{equation}
Clearly, at $\theta=\pi$, the angle $\varphi$ is ill defined and so is the
polarisation angle $\alpha$ (Figure~\ref{plane_wave}), but the difference
$\varphi-\alpha$ is meaningful. The maximum voltage and current occur at $\alpha=\varphi$
or $\pi-\varphi$, representing two opposite polarisations. Choosing $\alpha=\varphi$
the polarisation in Figure~\ref{plane_wave} becomes
$\boldsymbol{\widehat{\theta}}\cos\varphi+\boldsymbol{\widehat{\varphi}}\sin\varphi$,
which equals $-\widehat{x}$ as shown in Figure~\ref{left_incidence} and using $V^+=1$V
Eqs.~(\ref{voltage_left_incidence_gen}) and (\ref{current_left_incidence_gen}) become
\begin{equation}
V(z)=je^{-jkL}\sin[k(L-z)]
\label{voltage_left_incidence}
\end{equation}
\begin{equation}
Z_0I(z)=-je^{-jkL}\sin[k(L-z)],
\label{current_left_incidence}
\end{equation}
which behave oscillatory according to the frequency, with 0 value at right termination at $z=L$.
We remark that for this plane wave $E_z=H_z=0$ therefore the voltage and current can be measured
on any paths (see Figures~\ref{parallel_cylinders_Hz} and \ref{parallel_cylinders_Ez}).

\begin{figure}[!tbh]
\includegraphics[width=9cm]{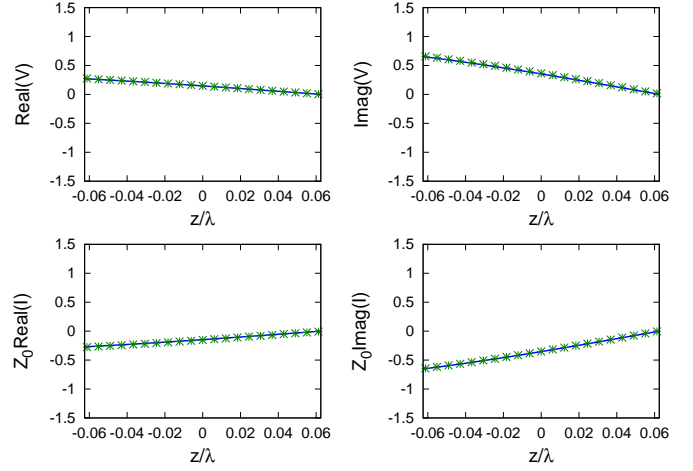}
\caption{Real and imaginary parts of the voltage $V(z)$ and normalized current $Z_0I(z)$ for the plane
wave incidence shown in Figure~\ref{left_incidence}, at frequency
30MHz or $L/\lambda=1/16$. The continuous line is the analytic solution in Eqs.~(\ref{voltage_left_incidence})
and (\ref{current_left_incidence}) and the stars are the ANSYS simulation results.}
\label{Ex_th_pi_30M}
\end{figure}

\begin{figure}[!tbh]
\includegraphics[width=9cm]{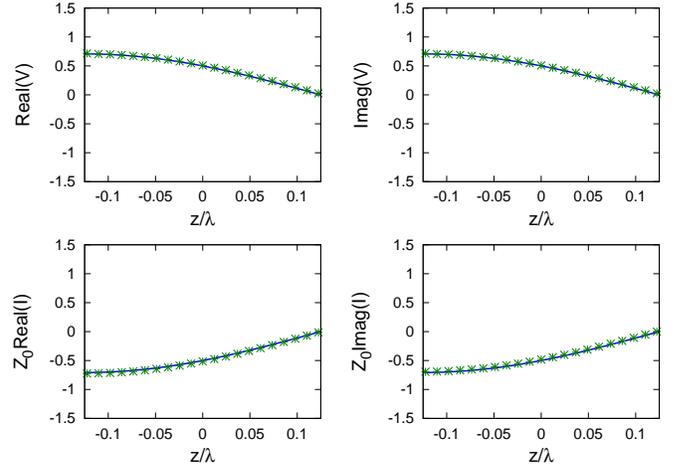}
\caption{Same as Figure~\ref{Ex_th_pi_30M}, for frequency 60MHz or $L/\lambda=1/8$.}
\label{Ex_th_pi_60M}
\end{figure}

\begin{figure}[!tbh]
\includegraphics[width=9cm]{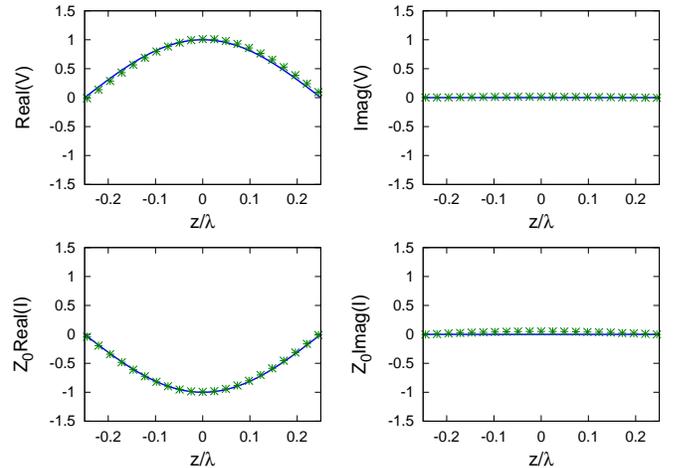}
\caption{Same as Figure~\ref{Ex_th_pi_30M}, for frequency 120MHz or $L/\lambda=1/4$.}
\label{Ex_th_pi_120M}
\end{figure}

Figures~\ref{Ex_th_pi_30M}-\ref{Ex_th_pi_120M} show the voltage and normalized current for the $-\mathbf{\widehat{x}}$ polarised
plane wave from $\theta=\pi$ in Figure~\ref{left_incidence} for frequencies 30, 60 and 120MHz, or $L/\lambda=1/16$,
1/8 and 1/4, respectively.
The voltage at the left termination at $z=-L$ is $V(-L)=je^{-jkL}\sin(2kL)$, resulting in
$0.27+j0.65$, $(1+j)/\sqrt{2}$ and 0 for the cases $L/\lambda=1/16$, 1/8 and 1/4, respectively.
The normalized currents $Z_0I(-L)$ gets the minus of the above values, satisfying the termination
condition $V(-L)=-Z_0I(-L)$.

\begin{figure}[!tbh]
\includegraphics[width=9cm]{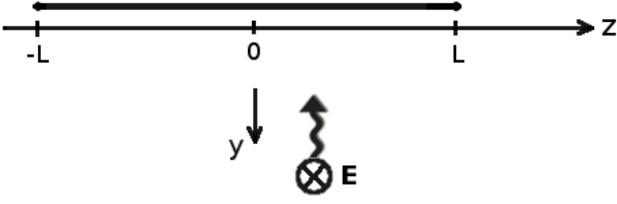}
\caption{Matched TL illuminated by a $-\mathbf{\widehat{x}}$ polarised plane wave from $(\theta=\pi/2,\varphi=\pi/2)$
(as also shown in Figure~\ref{parallel_cylinders_Hz}). The view is from the positive $x$ axis direction,
so that only the ``upper positive'' conductor is seen.}
\label{front_incidence}
\end{figure}

In the next example we use a plane wave hitting from $\theta=\varphi=\pi/2$ with
phase $e^{jky}$ as shown in Figure~\ref{front_incidence}.
For this direction Eqs.~(\ref{V_tot}) and (\ref{I_tot}) (normalized by $Z_0$) reduce to
\begin{equation}
V(z)=V^+\sin\alpha[1-e^{-jkL}\cos(kz)]
\label{voltage_front_incidence_gen}
\end{equation}
\begin{equation}
Z_0I(z)=jV^+\sin\alpha \,e^{-jkL}\sin(kz),
\label{current_front_incidence_gen}
\end{equation}
The maximum is for $\alpha=\pm\pi/2$, so using $\alpha=\pi/2$, results according to
Figure~\ref{plane_wave} in a $\boldsymbol{\widehat{\varphi}}$ polarisation which is
$-\widehat{x}$ at $\theta=\varphi=\pi/2$, as shown in Figure~\ref{front_incidence}.
We note that this is the same plane wave shown in Figure~\ref{parallel_cylinders_Hz}, so that here
the path on which one measures the voltage matters and must be path $A$ in Figure~\ref{parallel_cylinders_Hz}.

Using $V^+=1$V and $\alpha=\pi/2$ in Eqs.~(\ref{voltage_front_incidence_gen}) and (\ref{current_front_incidence_gen})
result in
\begin{equation}
V(z)=1-e^{-jkL}\cos(kz)
\label{voltage_front_incidence}
\end{equation}
\begin{equation}
Z_0I(z)=je^{-jkL}\sin(kz),
\label{current_front_incidence}
\end{equation}

\begin{figure}[!tbh]
\includegraphics[width=9cm]{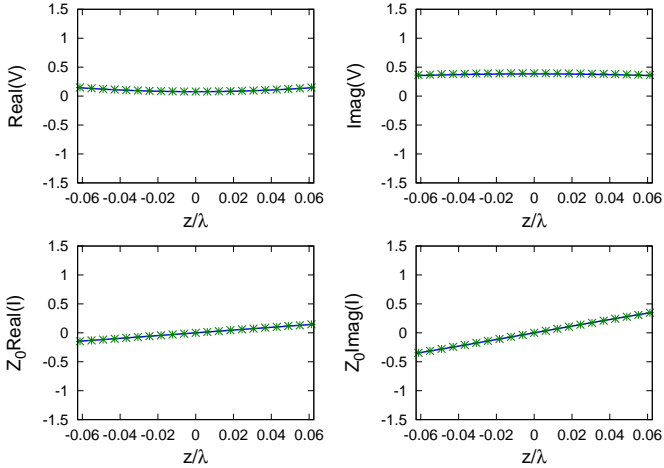}
\caption{Real and imaginary parts of the voltage $V(z)$ and normalized current $Z_0I(z)$ for the plane
wave incidence shown in Figure~\ref{front_incidence}, at frequency
30MHz or $L/\lambda=1/16$. The continuous line is the analytic solution in Eqs.~(\ref{voltage_front_incidence})
and (\ref{current_front_incidence}) and the stars are the ANSYS simulation results.}
\label{Ex_th_pi_over_2_phi_over_2_30M}
\end{figure}

\begin{figure}[!tbh]
\includegraphics[width=9cm]{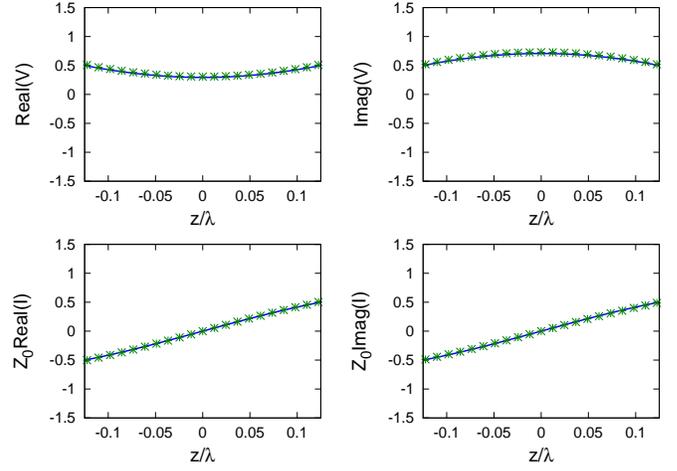}
\caption{Same as Figure~\ref{Ex_th_pi_over_2_phi_over_2_30M}, for frequency 60MHz or $L/\lambda=1/8$.}
\label{Ex_th_pi_over_2_phi_over_2_60M}
\end{figure}

\begin{figure}[!tbh]
\includegraphics[width=9cm]{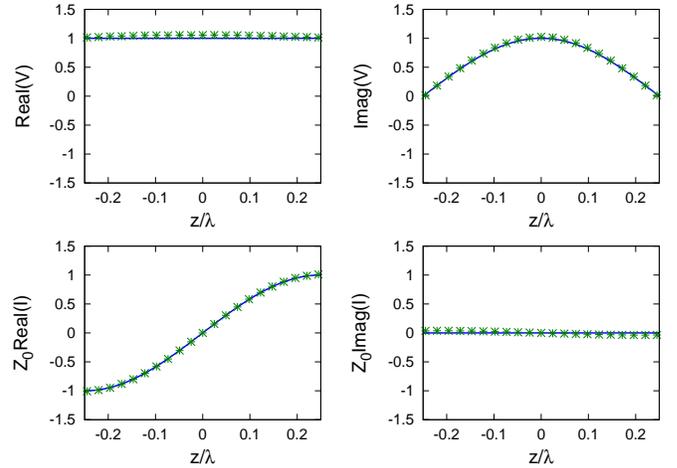}
\caption{Same as Figure~\ref{Ex_th_pi_over_2_phi_over_2_30M}, for frequency 120MHz or $L/\lambda=1/4$.}
\label{Ex_th_pi_over_2_phi_over_2_120M}
\end{figure}

Figures~\ref{Ex_th_pi_over_2_phi_over_2_30M}-\ref{Ex_th_pi_over_2_phi_over_2_120M} show the voltage and normalized
current for the $-\widehat{x}$ polarised plane wave from $\theta=\pi/2$ and $\varphi=\pi/2$
in Figure~\ref{left_incidence} for frequencies 30, 60 and 120MHz, or $L/\lambda=1/16$,
1/8 and 1/4, respectively.

As expected, the voltage is an even function of $z$ and the current an odd function of $z$. Specifically,
the voltages at the terminations at $z=-L$ and $L$ are $0.146+j0.354$, $0.5+j0.5$ and 1
for the cases $L/\lambda=1/16$, 1/8 and 1/4, respectively. The normalized currents have the same
values at $z=L$, satisfying the termination condition $V(L)=Z_0I(L)$, and minus the above values
at $z=-L$, satisfying $V(-L)=-Z_0I(-L)$.

The voltage in the middle of the TL at $z=0$ are 0.076+j0.383, $1-1/\sqrt{2}+j/\sqrt{2}$ and 1+j
for the cases $L/\lambda=1/16$, 1/8 and 1/4, respectively, and the current at $z=0$ is of course 0.

\begin{figure}[!tbh]
\includegraphics[width=9cm]{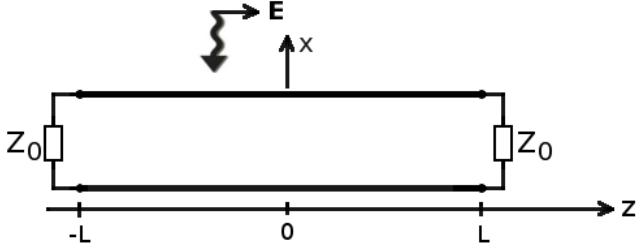}
\caption{Matched TL illuminated by a $\mathbf{\widehat{z}}$ polarised plane wave from $(\theta=\pi/2,\varphi=0)$
(as also shown in Figure~\ref{parallel_cylinders_Ez}).}
\label{up_incidence}
\end{figure}

In the next example we use a plane wave incident from $\theta=\pi/2$ and
$\varphi=0$, with phase $e^{jkx}$ as shown in Figure~\ref{up_incidence}.
For this direction Eqs.~(\ref{V_tot}) and (\ref{I_tot}) (normalized by $Z_0$) reduce to
\begin{equation}
V(z)=-jV^+\cos\alpha\, e^{-jkL}\sin(kz)
\label{voltage_up_incidence_gen}
\end{equation}
\begin{equation}
Z_0I(z)=-V^+\cos\alpha[1-e^{-jkL}\cos(kz)].
\label{current_up_incidence_gen}
\end{equation}
Here the maximum is obtained for $\alpha=0$ or $\pi$. Using $\alpha=\pi$ defines according
to Figure~\ref{plane_wave} a $-\boldsymbol{\widehat{\theta}}$ polarisation which equals
$\widehat{z}$ at $\theta=\pi/2$ and $\varphi=0$, as shown in Figure~\ref{up_incidence}.
This is the same plane wave
mentioned in Figure~\ref{parallel_cylinders_Ez}, which requires current measurement by integrating
on the tight trajectories $A$ and $B$ in Figure~\ref{parallel_cylinders_Ez}, using Eq.~(\ref{I_simul}) to determine the
differential current.

Using $V^+=1$V and $\alpha=\pi$ in Eqs.~(\ref{voltage_front_incidence_gen}) and (\ref{current_front_incidence_gen})
result in
\begin{equation}
V(z)=je^{-jkL}\sin(kz)
\label{voltage_up_incidence}
\end{equation}
\begin{equation}
Z_0I(z)=1-e^{-jkL}\cos(kz),
\label{current_up_incidence}
\end{equation}

\begin{figure}[!tbh]
\includegraphics[width=9cm]{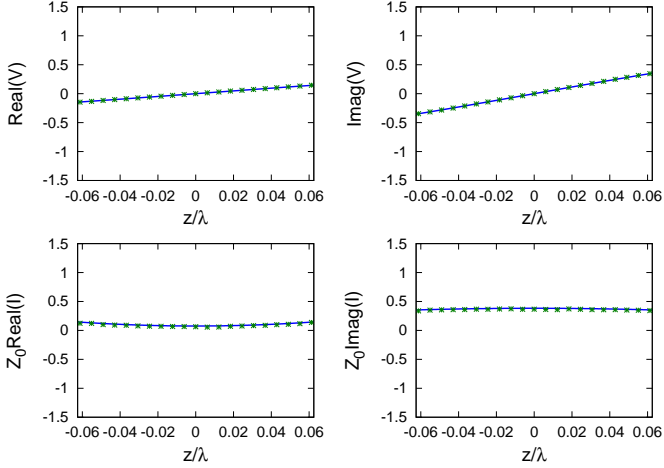}
\caption{Real and imaginary parts of the voltage $V(z)$ and normalized current $Z_0I(z)$ for the plane
wave incidence shown in Figure~\ref{up_incidence}, at frequency
30MHz or $L/\lambda=1/16$. The continuous line is the analytic solution in Eqs.~(\ref{voltage_up_incidence})
and (\ref{current_up_incidence}) and the stars are the ANSYS simulation results.}
\label{Ez_th_pi_over_2_phi_0_30M}
\end{figure}

\begin{figure}[!tbh]
\includegraphics[width=9cm]{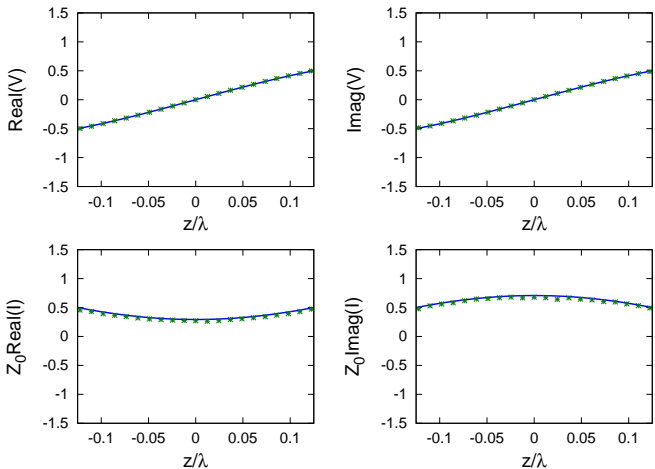}
\caption{Same as Figure~\ref{Ez_th_pi_over_2_phi_0_30M}, for frequency 60MHz or $L/\lambda=1/8$.}
\label{Ez_th_pi_over_2_phi_0_60M}
\end{figure}

\begin{figure}[!tbh]
\includegraphics[width=9cm]{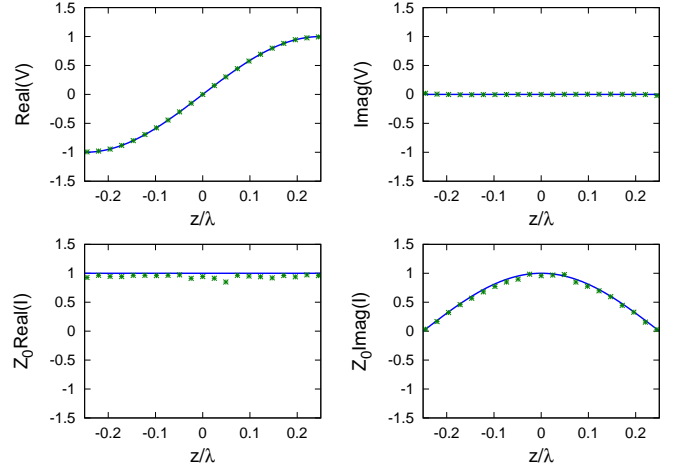}
\caption{Same as Figure~\ref{Ez_th_pi_over_2_phi_0_30M}, for frequency 120MHz or $L/\lambda=1/4$.}
\label{Ez_th_pi_over_2_phi_0_120M}
\end{figure}

Figures~\ref{Ez_th_pi_over_2_phi_0_30M}-\ref{Ez_th_pi_over_2_phi_0_120M} show the voltage and normalized
current for the $\mathbf{\widehat{z}}$ polarised plane wave from $\theta=\pi/2$ and $\varphi=0$
in Figure~\ref{up_incidence} for frequencies 30, 60 and 120MHz, or $L/\lambda=1/16$,
1/8 and 1/4, respectively.

As expected, the voltage is an odd function of $z$ and the current an even function of $z$. 
The voltage developed for the incident field in Figure~\ref{up_incidence} (Eq.~\ref{voltage_up_incidence})
has the value of the normalized current developed for the incident field in Figure~\ref{front_incidence}
(Eq.~\ref{current_front_incidence}) and viceversa.

\subsubsection{Non matched transmission line}
\label{non_matched}

We compare here several unmatched cases for the $-\widehat{x}$ polarised plane wave from $\theta=\pi/2$, and
$\varphi=\pi/2$, shown in Figure~\ref{front_incidence} for frequency 60~MHz. The matched solutions are given in
Eqs.~(\ref{voltage_front_incidence}) and (\ref{current_front_incidence}), and for this frequency ($kL=\pi/4$)
are:
\begin{equation}
V(z)=1-e^{-j\pi/4}\cos(kz)
\label{voltage_front_incidence_60M}
\end{equation}
\begin{equation}
Z_0I(z)=je^{-j\pi/4}\sin(kz),
\label{current_front_incidence_60M}
\end{equation}
as shown in Figure~\ref{Ex_th_pi_over_2_phi_over_2_60M}.

We generalize them for non matched cases using Eqs~(\ref{V_NM}) and (\ref{I_NM}), obtaining the correction
terms due to non matching (Eqs.~(\ref{Delta_V}) and (\ref{Delta_I}))
\begin{equation}
\Delta V(z)= \frac{\Gamma_L e^{-jkz}+\Gamma_R e^{jkz}-2j\Gamma_L\Gamma_R \cos(kz)}{\sqrt{2}(1+\Gamma_R\Gamma_L)}
\label{Delta_V_60M}
\end{equation}
\begin{equation}
Z_0\Delta I(z)= \frac{\Gamma_L e^{-jkz}-\Gamma_R e^{jkz}-2\Gamma_L\Gamma_R \sin(kz)}{\sqrt{2}(1+\Gamma_R\Gamma_L)}
\label{Delta_I_60M}
\end{equation}

\begin{figure}[!tbh]
\includegraphics[width=9cm]{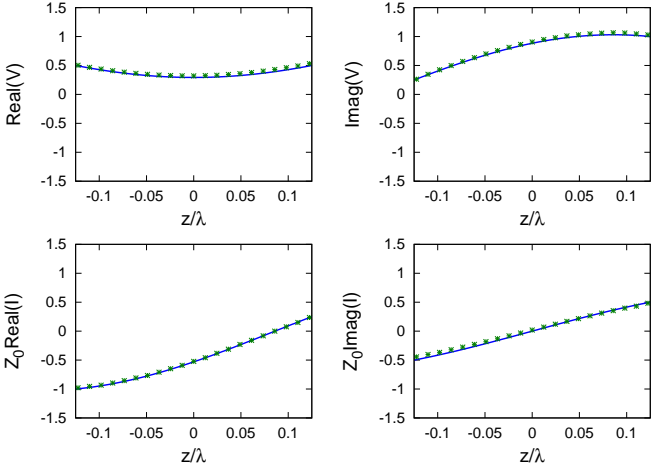}
\caption{Real and imaginary parts of the voltage $V(z)$ and normalized current $Z_0I(z)$ for the plane
wave incidence shown in Figure~\ref{front_incidence}, at frequency 60MHz (or $L/\lambda=1/16$), but for
loads $Z_L=Z_0/2$ and $Z_R=2Z_0$, or $\Gamma_L=-1/3$ and $\Gamma_R=1/3$. The continuous line is the analytic
solution, i.e. Eq.~(\ref{voltage_front_incidence_60M}) plus correction (\ref{Delta_V_60M})
and Eq.~(\ref{current_front_incidence_60M}) plus correction (\ref{Delta_I_60M}) and the stars
are the ANSYS simulation results.}
\label{Ex_th_pi_over_2_phi_over_2_60M_ZL_half_ZR_2}
\end{figure}

\begin{figure}[!tbh]
\includegraphics[width=9cm]{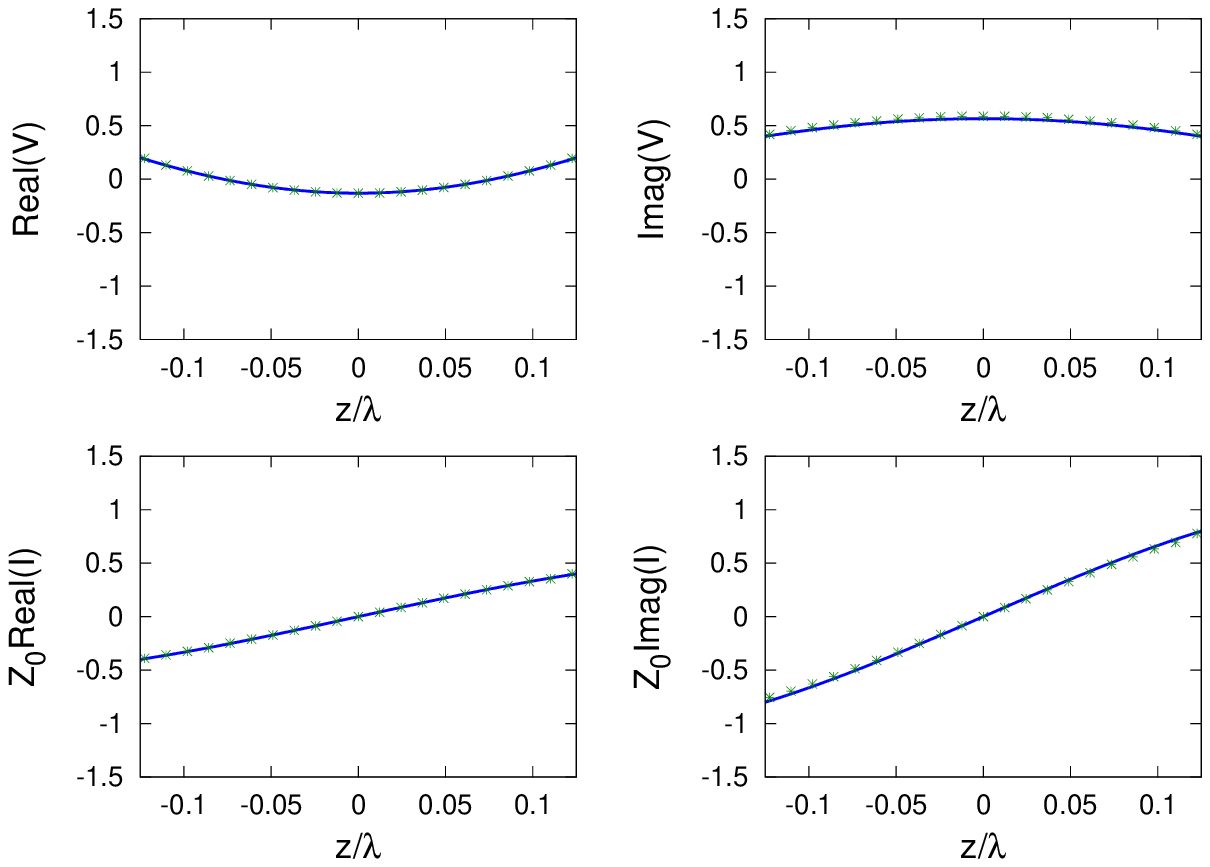}
\caption{Same as Figure~\ref{Ex_th_pi_over_2_phi_over_2_60M_ZL_half_ZR_2} for
$Z_L=Z_R=Z_0/2$ or $\Gamma_L=\Gamma_R=-1/3$.}
\label{Ex_th_pi_over_2_phi_over_2_60M_ZL_half_ZR_half}
\end{figure}

\begin{figure}[!tbh]
\includegraphics[width=9cm]{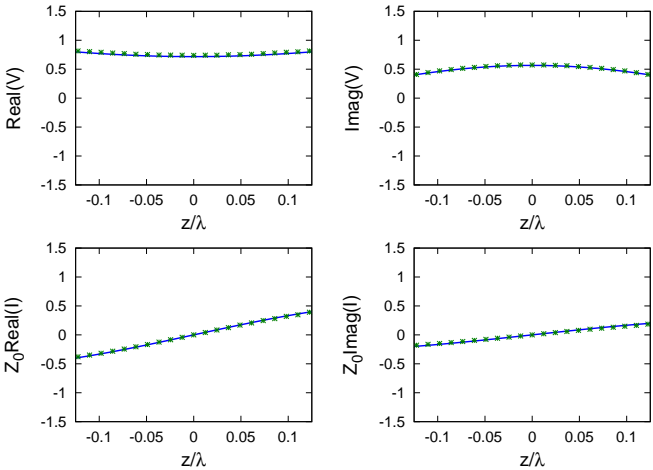}
\caption{Same as Figure~\ref{Ex_th_pi_over_2_phi_over_2_60M_ZL_half_ZR_2} for
$Z_L=Z_R=2Z_0$ or $\Gamma_L=\Gamma_R=1/3$.}
\label{Ex_th_pi_over_2_phi_over_2_60M_ZL_2_ZR_2}
\end{figure}

Figures~\ref{Ex_th_pi_over_2_phi_over_2_60M_ZL_half_ZR_2}-\ref{Ex_th_pi_over_2_phi_over_2_60M_ZL_2_ZR_2}
show the voltage and normalized current for the cases: $Z_L=Z_0/2$ and $Z_R=2Z_0$, $Z_L=Z_R=Z_0/2$
and $Z_L=Z_R=2Z_0$, respectively.

\subsection{Compatibility with previous works}
\label{validation_compatibility}

We show in this subsection that our analytic results are compatible with the
results obtained by other authors \cite{Taylor,Smith,Harrison,Paul,Agrawal}, hence
satisfy Eqs.~(\ref{telegraph1}) and (\ref{telegraph2}).

Using Eqs.~(\ref{Delta_V}), (\ref{Delta_I}) and (\ref{VI_minus_1_M}), it is easy to check that
\begin{equation}
\frac{d\Delta V}{dz}+jk (Z_0 \Delta I)=0\,\,\,\,\,\, ; \,\,\,\,\,\,Z_0\frac{d\Delta I}{dz}+jk \Delta V=0,
\label{sources_of_corrections_0}
\end{equation}
which is clear, because the induced sources depend only on the incident field and not on the loads.

Therefore, it is enough to show that Eqs.~(\ref{telegraph1}) and (\ref{telegraph2}) are satisfied
by the matched solution $V(z)$, $I(z)$ in Eqs.~(\ref{V_tot}), (\ref{I_tot}), for a general incident
plane wave.

First we need the phase of the incident plane wave (Eq.~(\ref{plane_wave_phase})) on the TL at $x\simeq 0$, $y=0$
and $-L\le z \le L$, which comes out
\begin{equation}
e^{jkz\cos\theta},
\label{plane_wave_phase_on_TL}
\end{equation}
so the incident $E_x$ and $H_y$ fields along the TL are
\begin{equation}
E_x(z)=E_xe^{jkz\cos\theta}\,\,\,\,\,\, ; \,\,\,\,\,\,H_y(z)=H_ye^{jkz\cos\theta},
\label{E_x_H_y_on_TL}
\end{equation}
where the values $E_x$ and $H_y$ are understood as the values at the origin.
Using the unit vectors identity
$\mathbf{\widehat{x}}=\mathbf{\widehat{r}}\sin\theta\cos\varphi+\boldsymbol{\widehat{\theta}}\cos\theta\cos\varphi-\boldsymbol{\widehat{\varphi}}\sin\varphi$, given the incident E field has only $\theta$ and $\varphi$ components, the $x$ component of the electric field (at origin) is
\begin{align}
E_x=&E_{\theta}\cos\theta\cos\varphi - E_{\varphi}\sin\varphi= \notag \\
    &E_0(\cos\alpha\cos\theta\cos\varphi-\sin\alpha\sin\varphi)
\label{E_x}
\end{align}
where for the second expression we used Eq.~(\ref{E_plane_wave_components}). Using
Eqs.~(\ref{Voltage_E_field_equivalence}), (\ref{E_x_H_y_on_TL}) and (\ref{E_x}),
$Z_0I_s$ in Eq.~(\ref{Is}) comes out
\begin{equation}
Z_0I_s=jk V^+(\sin\alpha\sin\varphi-\cos\alpha\cos\theta\cos\varphi)e^{jkz\cos\theta},
\label{Is_detailed}
\end{equation}
To derive the $H_y$ component of the incident field we note that the H components of the incident
plane wave are:
\begin{equation}
H_{\theta}=\frac{E_{\varphi}}{\eta_0}\,\,\,\,\,\, ; \,\,\,\,\,\,H_{\varphi}=-\frac{E_{\theta}}{\eta_0}
\label{H_components}
\end{equation}
Using the unit vectors identity
$\mathbf{\widehat{y}}=\mathbf{\widehat{r}}\sin\theta\sin\varphi+\boldsymbol{\widehat{\theta}}\cos\theta\sin\varphi+\boldsymbol{\widehat{\varphi}}\cos\varphi$, given the incident H field has only $\theta$ and $\varphi$ components,
the $y$ component of H (at origin) is
\begin{align}
H_y=&H_{\theta}\cos\theta\sin\varphi + H_{\varphi}\cos\varphi= \notag \\
    &\frac{1}{\eta_0}[E_{\varphi}\cos\theta\sin\varphi - E_{\theta}\cos\varphi],
\label{H_y}
\end{align}
where in the second expression we used Eq.~(\ref{H_components}). Using Eqs.~(\ref{E_plane_wave_components}) and 
(\ref{Voltage_E_field_equivalence}), this can be rewritten as
\begin{align}
\eta_0H_y=V^+[\sin\alpha\cos\theta\sin\varphi - \cos\alpha\cos\varphi],
\label{H_y_1}
\end{align}
Using Eqs.~(\ref{E_x_H_y_on_TL}), $V_s$ in Eq.~(\ref{Vs}) comes out
\begin{equation}
V_s= jk V^+[\sin\alpha\cos\theta\sin\varphi - \cos\alpha\cos\varphi]e^{jkz\cos\theta},
\label{Vs_detailed}
\end{equation}

Here it is left to show that Eqs.~(\ref{Is_detailed}) and (\ref{Vs_detailed}) equal to the LHS of Eqs.~(\ref{telegraph2})
and (\ref{telegraph1}) respectively, applied on the matched solution $V(z)$, $I(z)$ given in Eqs.~(\ref{V_tot}), (\ref{I_tot}).
We start with two intermediate results for the functions $f_1(z)$ and $f_2(z)$ defined in Eq.~(\ref{f_12}):
\begin{align}
&df_1/dz+jkf_1=k\cos^2(\theta/2)e^{jkz\cos\theta} \notag \\
&df_2/dz-jkf_2=-k\sin^2(\theta/2)e^{jkz\cos\theta},
\label{fprime_jk_12}
\end{align}
and obtain
\begin{align}
\frac{dV}{dz}+jk (Z_0I)&=-jkV^+[\cos^2(\theta/2)\cos(\varphi+\alpha)+ \notag \\
                       &\sin^2(\theta/2)\cos(\varphi-\alpha)]e^{jkz\cos\theta}
\label{telegraph1_detailed}
\end{align}
\begin{align}
Z_0\frac{dI}{dz}+jk V&=jkV^+[\sin^2(\theta/2)\cos(\varphi-\alpha)- \notag \\
                     &\cos^2(\theta/2)\cos(\varphi+\alpha)]e^{jkz\cos\theta}
\label{telegraph2_detailed}
\end{align}
Using simple trigonometric identities one finds that results (\ref{telegraph1_detailed}) and (\ref{telegraph2_detailed}) are identical
to Eqs.~(\ref{Vs_detailed}) and (\ref{Is_detailed}), respectively. 

\section{The power transferred to the loads}
\label{power}

We calculate here the power transferred to the loads for the general case of loads $Z_L$
and $Z_R$ as shown in Figures~\ref{config} and \ref{twin_lead_equivalent}. We need first
the voltage and current for the {\it matched at both teminations} case at the terminations ($z=\pm L$).
From Eqs.~(\ref{V_tot}) and (\ref{I_tot}) we obtain
\begin{equation}
V(-L)=jV^+ e^{-jkL}\sin[2kL\sin^2(\theta/2)]\cos(\varphi-\alpha)
\label{V_minus_L}
\end{equation}
\begin{equation}
V(L)=-jV^+ e^{-jkL}\sin[2kL\cos^2(\theta/2)]\cos(\varphi+\alpha)
\label{V_plus_L}
\end{equation}
and
\begin{equation}
Z_0I(-L)=-V(-L) \,\,\,\,\,\,\,;\,\,\,\,\,\,\, Z_0I(L)=V(L)
\label{I_minus_L_plus_L}
\end{equation}
satisfy the boundary conditions. From Eqs.~(\ref{V_minus_L})-(\ref{I_minus_L_plus_L}), the powers on the ``left''
and ``right'' loads for the {\it matched at both teminations} case, are
\begin{align}
P(-L)\equiv &-V(-L)I^*(-L)= \notag \\
          &\frac{|V^+|^2}{Z_0} \sin^2[2kL\sin^2(\theta/2)]\cos^2(\varphi-\alpha)
\label{P_matched_minus_L}
\end{align}
\begin{align}
P(L)\equiv &V(L)I^*(L)= \notag \\
          &\frac{|V^+|^2}{Z_0} \sin^2[2kL\cos^2(\theta/2)]\cos^2(\varphi+\alpha)
\label{P_matched_plus_L}
\end{align}

We now generalise the powers for a TL with any loads. We set $z=\pm L$ in Eqs.~(\ref{V_NM}),
(\ref{I_NM}), or simpler, use directly Eqs.~(\ref{V_NM_port_1}), (\ref{V_NM_port_M}), (\ref{I_NM_port_1})
and (\ref{I_NM_port_M}), obtaining
\begin{equation}
V_{\text{NM}}(-L)=(1+\Gamma_L)\frac{\Gamma_R e^{-j2kL}V(L)+V(-L)}{1-\Gamma_L\Gamma_R e^{-j4kL}}
\label{V_NM_minus_L}
\end{equation}
\begin{equation}
V_{\text{NM}}(L)=(1+\Gamma_R)\frac{\Gamma_L e^{-j2kL}V(-L)+V(L)}{1-\Gamma_L\Gamma_R e^{-j4kL}}
\label{V_NM_plus_L}
\end{equation}
and
\begin{equation}
Z_LI_{\text{NM}}(-L)=-V_{\text{NM}}(-L) \,\,\,\,\,\,;\,\,\,\,\,\, Z_RI_{\text{NM}}(L)=V_{\text{NM}}(L)
\label{I_NM_minus_L_plus_L}
\end{equation}
satisfy the boundary conditions.

Now we express the power on the ``left'' load $Z_L$ by $P_{\text{NM}}(-L)=Re\{-V_{\text{NM}}(-L)I^*_{\text{NM}}(-L)\}$
and on the ``right'' load $Z_R$ by
$P_{\text{NM}}(L)=Re\{V_{\text{NM}}(L)I^*_{\text{NM}}(L)\}$. Using Eqs.~(\ref{V_NM_minus_L})-(\ref{I_NM_minus_L_plus_L})
one obtains
\begin{align}
P_{\text{NM}}(-L)=&\frac{1-|\Gamma_L|^2}{|1-\Gamma_L\Gamma_R e^{-j4kL}|^2}[P(-L)+|\Gamma_R|^2P(L)-  \notag \\
                  &2 P_{\text{mix}} Re\{\Gamma_R e^{-j2kL}\}]
\label{P_NM_minus_L}
\end{align}
\begin{align}
P_{\text{NM}}(L)=&\frac{1-|\Gamma_R|^2}{|1-\Gamma_L\Gamma_R e^{-j4kL}|^2}[P(L)+|\Gamma_L|^2P(-L)-  \notag \\
                  &2 P_{\text{mix}} Re\{\Gamma_L e^{-j2kL}\}].
\label{P_NM_plus_L}
\end{align}
The generalised results for the powers are expressed in terms of the matched-TL powers $P(-L)$ and $P(L)$
(given in Eqs.~(\ref{P_matched_minus_L})-(\ref{P_matched_plus_L})) plus an additional ``mixed'' term $P_{\text{mix}}$
given by
\begin{align}
P_{\text{mix}}\equiv &V(L)I^*(-L)\equiv -V(-L)I^*(L)= \notag \\
          &\frac{|V^+|^2}{Z_0} \sin[2kL\sin^2(\theta/2)]\sin[2kL\cos^2(\theta/2)]  \notag \\
          &\cos(\varphi-\alpha)\cos(\varphi+\alpha)
\label{P_mix}
\end{align}

Those results may be understood from the radiation properties of the TL in \cite{full_model_arxiv}. We
remark $P(-L)$ is proportional to $\sin^2[2kL\sin^2(\theta/2)]$, according to the radiation
pattern of a TL carrying a forward wave only and maximised for the polarisation radiated in this case: 
\begin{equation}
\mathbf{\widehat{p}}^+(\theta,\varphi)=\boldsymbol{\widehat{\theta}}\cos\varphi+\boldsymbol{\widehat{\varphi}}\sin\varphi \,\,\,\,
\Rightarrow \,\,\,\, \alpha=\varphi,
\label{p_plus}
\end{equation}
see Section~2.1 in \cite{full_model_arxiv}. Also, we see $P(-L)=0$ for a polarisation orthogonal to
$\mathbf{\widehat{p}}^+$. This suggests that $P(-L)$, i.e. the power into the left load for the matched on both
sides TL is closely related to transmitting properties of a source at the left termination, issuing a forward wave.

Similarly, $P(L)$ is proportional to $\sin^2[2kL\cos^2(\theta/2)]$,
according to the radiation pattern of a TL carrying a backward wave only and maximised for the polarisation
radiated in this case:
\begin{equation}
\mathbf{\widehat{p}}^-(\theta,\varphi)=\boldsymbol{\widehat{\theta}}\cos\varphi-\boldsymbol{\widehat{\varphi}}\sin\varphi  \,\,\,\,
\Rightarrow \,\,\,\, \alpha=-\varphi,
\label{p_minus}
\end{equation}
see Section~2.2 in \cite{full_model_arxiv}, and $P(L)=0$ for a polarisation orthogonal to
$\mathbf{\widehat{p}}^-$. This suggests that $P(L)$, i.e. the power into the right load for the matched on both
sides TL is closely related to transmitting properties of a source at the right termination, issuing a backward wave.

The general non matched case given in Eqs.~(\ref{P_NM_minus_L})-(\ref{P_NM_plus_L}) is affected by both radiation
patterns and polarisations, and their combination found in the mixed term in Eq.~(\ref{P_mix}). We remark that for
a polarisation orthogonal to either $\mathbf{\widehat{p}}^+$ or $\mathbf{\widehat{p}}^-$, $P_{\text{mix}}=0$
and either $P(-L)$ or $P(L)$ are 0, in which case the received powers are either according to 
$\sin^2[2kL\cos^2(\theta/2)]$ or $\sin^2[2kL\sin^2(\theta/2)]$.

The formal connection between the transmitting and receiving properties of TL is analysed the next section.

\section{Consistency between transmitting-receiving radiation properties}
\label{transmit_receive}

To analyse the connection between the results of receiving electromagnetic radiation shown
in this work and the results of transmitting (radiating) electromagnetic radiation in
\cite{full_model_arxiv}, we have to summarise below some results from \cite{full_model_arxiv}
and rewrite them in a convenient form.

\begin{figure}[!tbh]
\includegraphics[width=9cm]{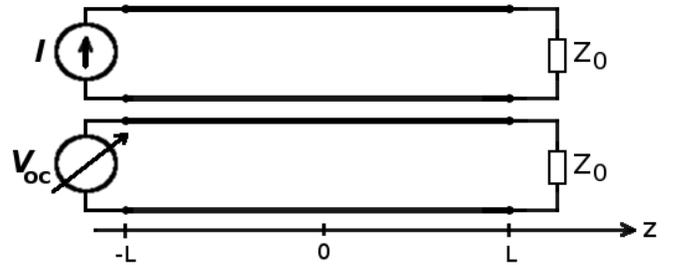}
\caption{Upper panel: radiating TL fed by a current source $I$ at $z=-L$, matched at $z=L$. Lower panel:
the same TL in receive mode, with a voltmeter at $z=-L$ measuring the open circuit voltage $V_{\text{oc}}$.}
\label{transm_TL_matched_at_right}
\end{figure}

The far electric field radiated by a TL carrying a forward wave only as in the upper panel of
Figure~\ref{transm_TL_matched_at_right},
is given in Section 2.1 of \cite{full_model_arxiv}
in terms of the value of $I^+$ in the middle of the TL at $z=0$. We rewrite it here in terms of
the current source $I$ in the standard form for a radiating antenna fed by a current
\begin{equation}
\mathbf{E}^+=jk\eta_0G(r)I\mathbf{l}^+_{\text{eff}},
\label{E_plus}
\end{equation}
where $\mathbf{l}^+_{\text{eff}}$ is the effective antenna length for the radiation of a TL carrying
a forward wave only, is given by
\begin{equation}
\mathbf{l}^+_{\text{eff}}=2jde^{-jkL}\sin[2kL\sin^2(\theta/2)]\mathbf{\widehat{p}}^+
\label{l_eff_plus}
\end{equation}
where the polarisation vector $\mathbf{\widehat{p}}^+$ is defined in Eq.~(\ref{p_plus}).

The radiation resistance for this case (given in Eq.(36) of \cite{full_model_arxiv}) is
\begin{equation}
r_{\text{rad}}=\frac{\eta_0}{2\pi} (kd)^2 \left[1-\sinc(4kL)\right],
\label{r_rad}
\end{equation}
which does not need a ``+'' superscript, being valid also for a backward wave only. The
radiation pattern (directivity) for a forward wave only is
\begin{equation}
D^+(\theta)=\frac{\eta_0k^2}{4\pi r_{\text{rad}}}|\mathbf{l}^+_{\text{eff}}|^2=2\frac{\sin^2[2kL\sin^2(\theta/2)]}{\left[1-\sinc(4kL)\right]},
\label{D_plus}
\end{equation}
as given in Section 2.1, Eq.~(13) in \cite{full_model_arxiv}. When dealing with radiation from TL, which is usually small relative to the
power carried by the TL, we may use the term ``radiation losses'' for the power lost to radiation. So in the upper panel of
Figure~\ref{transm_TL_matched_at_right}, the radiated power is $|I|^2r_{\text{rad}}$, and the power transmitted by the TL
is $|I|^2Z_0$, and clearly $r_{\text{rad}}\ll Z_0$. But in the context of the radiating properties we have to consider
the radiation efficiency, defined as the radiated power, divided by the total power into the antenna. In our
case this is
\begin{equation}
e_r=\frac{r_{\text{rad}}}{r_{\text{rad}}+Z_0}\simeq \frac{r_{\text{rad}}}{Z_0}.
\label{rad_eff}
\end{equation}
which is small, meaning that the (matched) TL is not an efficient antenna. Eq.~(\ref{rad_eff}) is valid for
both forward only or backward only wave, hence does not need a superscript. The antenna gain is the directivity
multiplied by the efficiency:
\begin{equation}
G^+(\theta)=D^+(\theta)e_r=\frac{(kd)^2\eta_0\sin^2[2kL\sin^2(\theta/2)]}{\pi Z_0},
\label{G_plus}
\end{equation}

Now, in receive mode, calculating the scalar product between the incident plane wave (Eq.~(\ref{E_plane_wave}))
and the effective length in Eq.~(\ref{l_eff_plus}) yields the open circuit voltage measured in the lower panel
of Figure~\ref{transm_TL_matched_at_right}
\begin{equation}
V_{\text{oc}}=\mathbf{E}\cdot\mathbf{l}^+_{\text{eff}} = 2jE_0\,d e^{-jkL}\sin[2kL\sin^2(\theta/2)]\cos(\varphi-\alpha),
\label{V_oc_E_dot_leff_plus}
\end{equation}
which is compared with the open end voltage obtained from the formalism developed in this work. So using
Eq.~(\ref{V_NM_minus_L}) with $\Gamma_R=0$ and $\Gamma_L=1$ (open) yield exactly the result
(\ref{V_oc_E_dot_leff_plus}) where we identify $V^+=E_0\,d$ according to Eq.~(\ref{Voltage_E_field_equivalence}).
Also, the absolute value of $V_{\text{oc}}$ is maximal for a matched polarisation, i.e. for $\alpha=\varphi$.

We replace now the voltmeter in the lower panel of Figure~\ref{transm_TL_matched_at_right} by a load $Z_L$
(left). Eq.~(\ref{P_NM_minus_L}), reduces for $\Gamma_R=0$ to $P_{\text{NM}}(-L)=(1-|\Gamma_L|^2)P(-L)$, so
maximum power is obtained for $\Gamma_L=0$ ($Z_L=Z_0$), yielding Eq.~(\ref{P_matched_minus_L}). Using the matched
polarisation ($\alpha=\varphi$) in (\ref{P_matched_minus_L}) results in the maximum received power:
\begin{equation}
P_{\text{rec}}=\frac{|V^+|^2}{Z_0} \sin^2[2kL\sin^2(\theta/2)].
\label{P_rec}
\end{equation}
Dividing this by the Poynting vector $S=E^2_0/\eta_0$, results in the
effective receiving cross section area  $A^+$ (for receiving into the left termination - the equivalent
of transmitting a forward wave)
\begin{equation}
A^+=\frac{P_{\text{rec}}}{S}=\frac{d^2 \eta_0}{Z_0} \sin^2[2kL\sin^2(\theta/2)].
\label{P_rec_over_S}
\end{equation}
which equals exactly to $\frac{\lambda^2}{4\pi}G^+$.

A similar analysis for the backward wave, using the current source at the $z=L$ termination
(pointing upward) and the matched load at $z=-L$ in Figure~\ref{transm_TL_matched_at_right}
yields the field radiated by a backward wave current:
\begin{equation}
\mathbf{E}^-=jk\eta_0G(r)I\mathbf{l}^-_{\text{eff}},
\label{E_minus}
\end{equation}
where $\mathbf{l}^-_{\text{eff}}$ is the effective antenna length for the radiation of a TL carrying
a backward wave only, is given by
\begin{equation}
\mathbf{l}^-_{\text{eff}}=-2jde^{-jkL}\sin[2kL\cos^2(\theta/2)]\mathbf{\widehat{p}}^-,
\label{l_eff_minus}
\end{equation}
and the polarisation vector $\mathbf{\widehat{p}}^-$ is defined in Eq.~(\ref{p_minus}).
For this case, the gain is the directivity function $D^-$ given in Eq.~(17) of \cite{full_model_arxiv}
multiplied by the efficiency in Eq.~(\ref{rad_eff}), yields
\begin{equation}
G^-(\theta)=\frac{(kd)^2\eta_0\sin^2[2kL\cos^2(\theta/2)]}{\pi Z_0},
\label{G_minus}
\end{equation}
The effective receiving cross section area $A^-$ comes out
\begin{equation}
A^-=\frac{d^2 \eta_0}{Z_0} \cos^2[2kL\cos^2(\theta/2)],
\label{A_minus}
\end{equation}
satisfying the relation $A^-=\frac{\lambda^2}{4\pi}G^-$.

\section{Conclusions}

We derived in this work the voltage and differential current developed on an ideal
two-conductors TEM transmission line (TL) of any small electric cross section, connected
to passive loads and hit by a monochromatic plane wave, as shown in Figure~\ref{config}.

For this derivation we used our knowledge on the radiation properties of TL \cite{full_model_arxiv}
to build S matrices which describe the radiating system and used the reciprocity to derive the current
and voltage induced on the TL. This methodology allowed us to derive first the voltage
and current on a matched at both teminations TL, yielding the relatively simple expressions
given in Eqs.~(\ref{V_tot}) and (\ref{I_tot}). The generalisation to any loads is then
obtained, using the S matrix.

We validated our analytic results in Section~\ref{validation_hfss} for both the matched on two
sides TL case and the non matched case for different plane wave incidences. We also showed 
in Section~\ref{validation_compatibility} that our analytic results are compatible with the
methodology used in previous works \cite{Taylor,Smith,Harrison,Paul,Agrawal}. The simplicity
of this proof, requiring the application of Eqs.~(\ref{telegraph1}) and (\ref{telegraph2})
{\it only} on the matched at both teminations solution (see Eq.~\ref{sources_of_corrections_0}), emphasises the
added value of this work.

In Section~\ref{power}, we derived the powers on the loads and showed the connection between those powers
and the radiating properties of the TL. For the matched at both teminations TL, the power on the ``left'' load
has the same spatial depencence on the incident plane wave direction as the radiation pattern of
a forward wave, and is maximised for the polarisation of a radiating forward wave
$\mathbf{\widehat{p}}^+$ (see Eq.~(\ref{p_plus})). Similarly, for the matched at both teminations
TL, the power on the ``right'' load has the same spatial depencence on the incident plane wave direction
as the radiation pattern of a backward wave, and is maximised for the polarisation $\mathbf{\widehat{p}}^-$
(see Eq.~(\ref{p_minus})). The general non matched case is affected by both radiation
patterns and polarisations, and their combination found in a mixed term.

In Section~\ref{transmit_receive} we showed that the formal relations between receiving cross section
area and antenna gain are satisfied for the TL.


%



\appendices
\renewcommand\thefigure{\thesection.\arabic{figure}}
\setcounter{figure}{0}
\renewcommand\theequation{\thesection.\arabic{equation}}
\setcounter{equation}{0}
\section{Far radiated E field and its normalization}
\label{normalization}

\begin{figure}[!tbh]
\includegraphics[width=9cm]{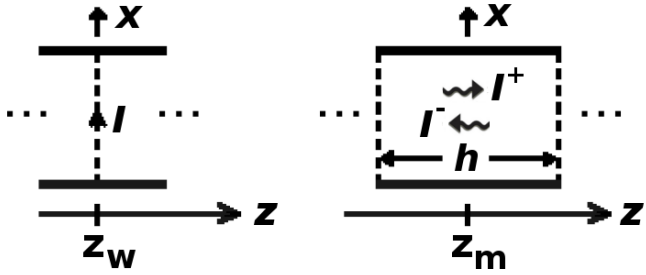}
\caption{Currents contributing to radiation. The left panel shows a $x$ directed current $I$ at location $z_w$
on the TL (which may represent a parallel port or a load), contributes to the far E field according to Eqs.
Eqs.(\ref{E_theta_x}) and (\ref{E_phi_x}). The right panel shows a TL segment of width $h$ carrying a forward
and/or backward current wave, $I^{\pm}$ evaluated at the middle location $z_m$, contributing to the far E field
according to Eq.~(\ref{E_theta_plus_minus_z}).}
\label{z_wire_segment}
\end{figure}

To derive the S matrices we need the contribution of currents shown in Figure~\ref{z_wire_segment}
to the far field. Based on 
\cite{full_model_arxiv}, a current element $Id$ in the $x$ direction (where $I$ is the current and
$d$ the separation distance in the twin lead representation), at $z=z_w$ (w=wire) contributes the following
E field in $\theta$ and $\varphi$ polarisations (see left panel of Figure~\ref{z_wire_segment}).
\begin{equation}
E_{\theta \,(x)}=-jkG(r)\eta_0 d I e^{jkz_w\cos\theta}\cos\theta\cos\varphi
\label{E_theta_x}
\end{equation}
\begin{equation}
E_{\varphi \,(x)}= jkG(r)\eta_0 d I e^{jkz_w\cos\theta}\sin\varphi,
\label{E_phi_x}
\end{equation}
where the subscript $(x)$ shows that this is a $x$ directed current contribution.

A TL section of lenght $h$ around $z=z_m$ (m=middle) carrying a forward and/or backward current contributes
the following far $\theta$ polarised E fields (see right panel of Figure~\ref{z_wire_segment}).
\begin{align}
E_{\theta \,(z)}=&-2 kG(r)\eta_0 d I^{\pm}(z_m)e^{jkz_m\cos\theta}(1\pm\cos\theta)\cos\varphi \notag \\
           &\sin[kh(1\mp\cos\theta)/2]
\label{E_theta_plus_minus_z}
\end{align}
where the upper and lower signs are for forward and backward waves respectively, and the values used for the
forward or backward currents are in the middle of the line at $z_m$, both currents defined in the $+z$ direction.
The subscript $(z)$ shows that this is a $z$ directed current contribution.

\begin{figure}[!tbh]
\includegraphics[width=8cm]{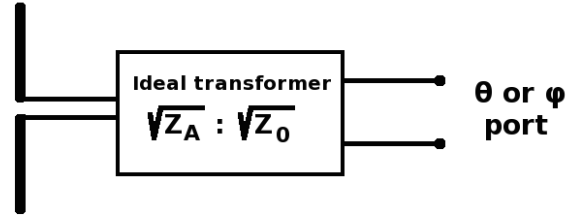}
\caption{The antenna model assumes a real antenna impedance $Z_A$ matched by an ideal transformer to the
$\theta$ or $\varphi$ port of impedance $Z_0$.}
\label{antenna_model}
\end{figure}

In the final result, $E_{\theta}=E_{\theta \,(x)}+E_{\theta \,(z)}$ and $E_{\varphi}=E_{\varphi \,(x)}$. To
work out the S matrices (Section~\ref{Voltage_S_matrix} and Appendix~\ref{Current_S_matrix}), one needs to scale
the radiated $E_{\theta}$ and $E_{\varphi}$ fields to outgoing voltages $V_{\theta}^-$ and $V_{\varphi}^-$
at the ports $\theta$ and $\varphi$ (Section~\ref{Voltage_S_matrix}), or to outgoing currents
$I_{\theta}^-$ and $I_{\varphi}^-$
(Appendix~\ref{Current_S_matrix}). Those ports have been defined for the impedance $Z_0$, which is the
charateristic impedance of the analysed TL. To make those ports matched, we use the antenna model
shown in Figure~\ref{antenna_model}.
The model assumes a real antenna impedance $Z_A$ matched via an ideal transformer to the 
$\theta$ or $\varphi$ port of impedance $Z_0$. Using the effective length of the
antenna $l_{\text{eff}}$ (for the given incidence direction), one expresses the open
circuit voltage on the antenna \cite{orfanidis}
\begin{equation}
V_{\text{oc}}=E\, l_{\text{eff}},
\label{V_oc}
\end{equation}
where $E$ is $E_{\theta}$ or $E_{\varphi}$. The $\theta$ or $\varphi$ port being matched, the antenna
``sees'' a matched load, so that the actual voltage on the antenna terminals is $V_{\text{oc}}/2$, and the
outgoing voltage at the port is
\begin{equation}
V^-=\frac{1}{2}E \, l_{\text{eff}} \sqrt{Z_0/Z_A}.
\label{V_minus}
\end{equation}
Now to activate the antenna in transmit mode, we feed the $\theta$ or $\varphi$ port by $V^+$,
obtaining the current feeding the antenna
\begin{equation}
I_A=V_A/Z_A=V^+/\sqrt{Z_AZ_0}
\label{I_A}
\end{equation}
Using the well known formula for the far E field radiated by a dipole \cite{orfanidis}
\begin{equation}
E_0=jk\eta_0G(r)I_A l_{\text{eff}}
\label{far_E_dipole}
\end{equation}
where $l_{\text{eff}}$ is the effective dipole length discussed before, into the same direction \cite{orfanidis}.
The field radiated from the antenna(s) has been named $E_0$ to distinguish it from the previously
discussed field travelling toward the antenna(s). Expressing it as function of $V^+$, we obtain
\begin{equation}
E_0=jk\eta_0G(r) l_{\text{eff}}V^+/\sqrt{Z_AZ_0}
\label{far_E_dipole_Vp}
\end{equation}
This field $E_0$ radiated from the far antenna(s), is the field incident on the TL (or one of its components),
mentioned in Figure~\ref{plane_wave}. We want to normalise it so that $E_0\,d=V^+$
(see Eq.~(\ref{Voltage_E_field_equivalence})), considering the field incident on the TL
as a plane wave. Imposing condition (\ref{Voltage_E_field_equivalence}) on
Eq.~(\ref{far_E_dipole_Vp}), results in
\begin{equation}
l_{\text{eff}}=\frac{\sqrt{Z_AZ_0}}{jk\eta_0G(r)d}
\label{l_eff_normalization}
\end{equation}

Given the far antennas are only tools to build the S matrices and from them infer the voltage
and current along the TL, the ``actual'' value of $l_{\text{eff}}$ is of no interest, so to express
the fields radiated by the TL $E_{\theta}$ and $E_{\varphi}$ as voltages, we set 
$l_{\text{eff}}$ from (\ref{l_eff_normalization}) into Eq.~(\ref{V_minus}), obtaining
\begin{equation}
V^-_{\theta \,\text{or}\, \varphi}=\frac{Z_0 E_{\theta\, \text{or}\, \varphi}}{2jk\eta_0G(r)d}
\label{scale_factor}
\end{equation}

For the case we need to express the fields radiated by the TL $E_{\theta}$ and $E_{\varphi}$ as
outgoing currents (Section~\ref{Current_S_matrix}), we use $I^-=-V^-/Z_0$, obtaining:
\begin{equation}
I^-_{\theta \,\text{or}\, \varphi}=-\frac{E_{\theta\, \text{or}\, \varphi}}{2jk\eta_0G(r)d}
\label{scale_factor_current}
\end{equation}

\renewcommand\thefigure{\thesection.\arabic{figure}}
\setcounter{figure}{0}
\renewcommand\theequation{\thesection.\arabic{equation}}
\setcounter{equation}{0}
\section{Generalised scattering matrix}
\label{Generalised_S_matrix}

\begin{figure}[!tbh]
\includegraphics[width=8cm]{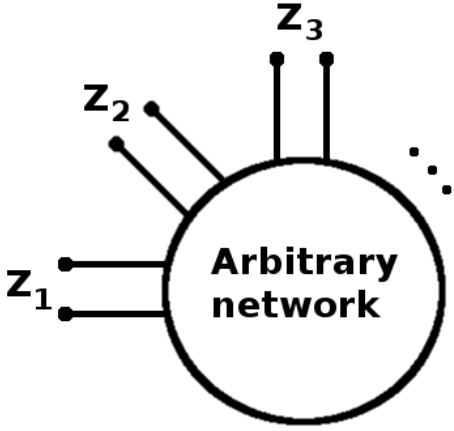}
\caption{An $N$ ports arbitrary network interfaced to transmission lines of characteristic impedances $Z_1$, $Z_2$, ... $Z_N$.}
\label{arbitrary_network}
\end{figure}

In this appendix we define the generalized scattering matrix and explain its reciprocity properties.
The generalized scattering matrix \cite{pozar} for an arbitrary network (as shown in Figure~\ref{arbitrary_network})
is defined by the following matrix equation
\begin{equation}
\mathbf{V}^- = \mathbf{S} \mathbf{V}^+,
\label{GSM}
\end{equation}
where $\mathbf{V}^{\pm}$ are column vectors for the incoming and outgoing voltage waves at the ports
of the network, where each port has its own characteristic impedance, as shown in Figure~\ref{arbitrary_network}.
The voltage and current waves at the ports satisfy
\begin{equation}
\mathbf{V}^+ = \mathbf{Z} \mathbf{I}^+ \,\,\,\,; \,\,\,\, \mathbf{V}^- = - \mathbf{Z} \mathbf{I}^-,
\label{V_p_V_m}
\end{equation}
where $\mathbf{Z}$ is a diagonal matrix of the characteristic impedances in Figure~\ref{arbitrary_network}:
\begin{equation}
Z_{ij}=Z_i\delta_{ij},
\label{Z_ij}
\end{equation}
so one can easily apply functions on them, as follows.

Normalising the voltages and currents at each port according to
\begin{equation}
\mathbf{a}^{\pm}\equiv\sqrt{\mathbf{Z}^{-1}}\mathbf{V}^{\pm} \,\,\,\,; \,\,\,\, \mathbf{b}^{\pm}\equiv\sqrt{\mathbf{Z}}\mathbf{I}^{\pm},
\label{a_b}
\end{equation}
and setting into Eq.~(\ref{GSM}), result in
\begin{equation}
\mathbf{a}^- =\sqrt{\mathbf{Z}^{-1}} \mathbf{S} \sqrt{\mathbf{Z}}\mathbf{a}^+,
\label{GSM_1}
\end{equation}
which define the (ordinary) scattering matrix, for which we use a lower case ``s''
\begin{equation}
\mathbf{s}\equiv \sqrt{\mathbf{Z}^{-1}} \mathbf{S} \sqrt{\mathbf{Z}}.
\label{OSM}
\end{equation}
It is known \cite{pozar,collin,orfanidis,ramo,jordan,balanis} that the reciprocity condition for the ordinary
scattering matrix is
\begin{equation}
s_{ij}=s_{ji}.
\label{OSM_reciprocity}
\end{equation}
Using this in Eq.~(\ref{OSM}) results in the reciprocity condition for the generalized scattering matrix
\begin{equation}
S_{ij}Z_j=S_{ji}Z_i
\label{GSM_reciprocity}
\end{equation}
Scattering matrices are usually defined for voltage waves, but in this work we need to define a
generalized scattering matrix for current waves as follows
\begin{equation}
\mathbf{I}^- = \mathbf{S}_I \mathbf{I}^+,
\label{GSM_I}
\end{equation}
and we name it $\mathbf{S}_I$ to distinguish it from the regular scattering matrix in (\ref{GSM}).
Using Eq.~(\ref{V_p_V_m}), one obtains
\begin{equation}
\mathbf{V}^- = -\mathbf{Z} \mathbf{S}_I \mathbf{Z}^{-1} \mathbf{V}^+.
\label{GSM_I_1}
\end{equation}
Comparing it with (\ref{GSM}) we get the relation
\begin{equation}
\mathbf{S} = -\mathbf{Z} \mathbf{S}_I \mathbf{Z}^{-1}.
\label{GSM_GSMI}
\end{equation}
Setting it in (\ref{OSM}) results in
\begin{equation}
\mathbf{s}= - \sqrt{\mathbf{Z}} \mathbf{S} \sqrt{\mathbf{Z}^{-1}},
\label{OSM_SI}
\end{equation}
and using (\ref{OSM_reciprocity}) we obtain
\begin{equation}
S_{I\,ij}Z_i=S_{I\,ji}Z_j
\label{GSM_I_reciprocity}
\end{equation}

\renewcommand\thefigure{\thesection.\arabic{figure}}
\setcounter{figure}{0}
\renewcommand\theequation{\thesection.\arabic{equation}}
\setcounter{equation}{0}
\section{Derivation of the current along the TL}
\label{Current_S_matrix}

We define a system of $M+2$ ports, the first $M$ ports are serial ports on the TL (shown in Figure~\ref{current_S})
and 2 additional ports are far antennas at polarisations $\boldsymbol{\widehat{\theta}}$ and $\boldsymbol{\widehat{\varphi}}$.

Our purpose is to determine the TL current for an incidence of a plane wave, we therefore calculate the
matrix $S$ defined by
\begin{equation}
\mathbf{I}^-=\mathbf{S} \mathbf{I}^+,
\label{S_for_currents}
\end{equation}
although this is not the usual definition of the scattering matrix (see Appendix~\ref{Generalised_S_matrix}).
For simplicity, we still call it $S$, but take into consideration that any reflection coefficient has an opposite sign. 

Feeding port $1<n<M$, located at $z_n$ with a forward (entering) current $I^+_n$ and terminating the other
TL ports $1\le i\le M$ by the impedances defined for those ports (i.e. $Z_0$ for ports 1 and $M$, and $Z_T$
for the middle ports) results in a forward wave from $z=z_n$ to $z=L$ and a backward wave from $z=z_n$
to $z=-L$, because the waves encounter at the intermediate ports a very small serial impedance $Z_T\to 0$.

The reflection coefficient at port $n$ is $S_{n,n}=\frac{Z_T-2Z_0}{Z_T+2Z_0}\simeq -1+\frac{Z_T}{Z_0}$, so the port
current $I_n=I^+_n(1+S_{nn})\simeq I^+_n\frac{Z_T}{Z_0}$. The current waves at ports $i\neq n$
give rise to outgoing currents at the ports given by:
\begin{equation}
I^-_{i\neq n}=I_ne^{-jk\Delta z|i-n|}=I^+_n\frac{Z_T}{Z_0}e^{-jk\Delta z|i-n|}.
\label{I_minus_i_neq_k}
\end{equation}
and for $i=n$
\begin{equation}
I^-_n=I^+_nS_{n,n}=I^+_n\left(-1+\frac{Z_T}{Z_0}\right)
\label{V_minus_kk_ser}
\end{equation}

This results in the following (partial) $n$ column of the S matrix
\begin{equation}
S_{1\le i\le M\,,\,1<n<M}=
\left\{
\begin{array}{l l}
\frac{Z_T}{Z_0}e^{-jk\Delta z|i-n|} &  \,\, i\neq n  \\
-1+\frac{Z_T}{Z_0}\simeq -1     &  \,\, i=n
\end{array}
\right.,
\label{partial_column_k_ser}
\end{equation}
see lower sign in Figure~\ref{matrix}. For column $n=1$ or $M$, the results are similar, only replace $\frac{Z_T}{Z_0}$ by 1, hence
\begin{equation}
S_{1\le i\le M\,,\,n=1,M}=
\left\{
\begin{array}{l l}
e^{-jk\Delta z|i-n|} &  \,\, i\neq n  \\
0                &  \,\, i=n
\end{array}
\right.
\label{partial_column_k_eq_1_or_M_ser}
\end{equation}

Eqs.~(\ref{partial_column_k_ser}) and (\ref{partial_column_k_eq_1_or_M_ser}) define the upper left square of
the S matrix, i.e. all the elements connecting the TL ports 1 to $M$, and now we calculate the two additional
elements belonging to ports $\theta$ and $\varphi$ (last two columns in Figure~\ref{matrix}), using
Eqs.~(\ref{E_theta_x})-(\ref{E_theta_plus_minus_z}) and scaling the fields with Eq.~(\ref{scale_factor_current}).

\begin{figure}[!tbh]
\includegraphics[width=9cm]{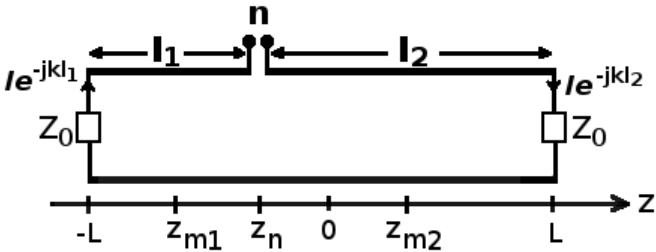}
\caption{Currents in the TL while feeding port $n$ with $I^+_n$. Defining $I\equiv I^+_n Z_T/Z_0$,
the $x$ directed currents are $Ie^{-jkl_1}$ in port 1 at $z=-L$ and $-Ie^{-jkl_2}$ in port $M$
at $z=L$. The $z$ directed current waves are the forward wave in the region $[z_n,L]$
and the backward current wave in the region $[-L,z_n]$, to be used in
Eq.~(\ref{E_theta_plus_minus_z}). For the forward wave we use the current at the middle point
$z_{m2}=(z_n+L)/2$, which is $Ie^{-jkl_2/2}$, while for the backward wave we use the current at $z_{m1}=(z_n-L)/2$
which us $Ie^{-jkl_1/2}$.}
\label{currents_configuration_ser}
\end{figure}

Feeding a {\it middle} port $1<n<M$ with the current $I^+_n$ and terminating all other ports with matched loads,
we calculate the far fields
given in Eqs.~(\ref{E_theta_x})-(\ref{E_theta_plus_minus_z}), scaled to the currents $I^-_{\theta}$ and $I^-_{\varphi}$
according to Eq.~(\ref{scale_factor_current}). As in the voltage calculation section, the value of $n$ is hidden
in the parameters $l_1$ and $l_2$ (Eq.~(\ref{l_12_k})), see Figure~\ref{currents_configuration_ser}.

We start with the contribution of the $x$ currents given in Eqs.~(\ref{E_theta_x}) and (\ref{E_phi_x}).
The $x$ directed currents are shown in Figure~\ref{currents_configuration_ser}.
Using Eqs.(\ref{E_theta_x}) for the two currents at $L$ and $-L$, scaling with (\ref{scale_factor_current}), we obtain
after some algebra:
\begin{equation}
I^-_{\theta \,(x)}=jI^+_n(Z_T/Z_0)[f_2-f_1]\cos\theta\cos\varphi,
\label{I_minus_theta_x_1}
\end{equation}
where $f_{1,2}$ are defined in Eq.~(\ref{f_12}). As evident from Eqs.(\ref{E_theta_x}) and (\ref{E_phi_x}),
$I^-_{\varphi \,(x)}$ is identical to the above, up
to replacing $\cos\theta\cos\varphi$ by $-\sin\varphi$, we therefore have
\begin{equation}
I^-_{\varphi \,(x)}=jI^+_n(Z_T/Z_0)[f_1-f_2]\sin\varphi
\label{I_minus_phi_x}
\end{equation}
Next we calculate the contributions of the forward current wave in the region $z=[z_n,L]$ and the backward current wave
in the region $z=[-L,z_n]$ (see Figure~\ref{currents_configuration_ser}), using the upper and lower signs in
Eq.~(\ref{E_theta_plus_minus_z}), respectively. For the forward wave we use the current in the middle point
$z_{m2}$, having the phase $e^{-jkl_2/2}$ for a TL length $h=l_2$, while for the backward wave we use the
current at $z_{m1}$, having the phase $e^{-jkl_1/2}$, for a TL length $h=l_1$, both currents are
defined in the $+z$ direction, hence taken with plus sign. Using $z_{m1,2}$ in Eq.~(\ref{z_m12}), after scaling with
(\ref{scale_factor_current}), we obtain:
\begin{equation}
I^-_{\theta \,(z)}=-2jI^+_n(Z_T/Z_0)[f_2\cos^2(\theta/2)+f_1\sin^2(\theta/2)]\cos\varphi
\label{I_minus_theta_z}
\end{equation}
Now we add the $x$ and $z$ directed currents contributions to $I^-_{\theta}=I^-_{\theta \,(x)}+I^-_{\theta \,(z)}$,
using $1-\cos(\theta)=2\sin^2(\theta/2)$ and $1+\cos(\theta)=2\cos^2(\theta/2)$, after some algebra we obtain:
\begin{equation}
I^-_{\theta}=-jI^+_n(Z_T/Z_0)[f_1+f_2]\cos\varphi
\label{I_minus_theta}
\end{equation}
$I^-_{\varphi}$ has only the contribution of the $x$ directed currents, given in Eq~(\ref{I_minus_phi_x}), so that
\begin{equation}
I^-_{\varphi}=I^-_{\varphi \,(x)}
\label{I_minus_phi}
\end{equation}
Results (\ref{I_minus_theta}) and (\ref{I_minus_phi}) define the $S_{\theta,n}$ and
$S_{\varphi,n}$ matrix elements respectively, for the columns $1<n<M$:
\begin{equation}
S_{\theta,\,1<n<M}=-j(Z_T/Z_0)[f_1+f_2]\cos\varphi
\label{S_theta_k_ser}
\end{equation}
\begin{equation}
S_{\varphi,\,1<n<M}=j(Z_T/Z_0)[f_1-f_2]\sin\varphi
\label{S_phi_k_ser}
\end{equation}
For column $n=1$ or $M$, the results are similar, only replace $\frac{Z_T}{Z_0}$ by 1, hence
\begin{equation}
S_{\theta,\,n=1,M}=-j[f_1+f_2]\cos\varphi
\label{S_theta_k_1_M_ser}
\end{equation}
\begin{equation}
S_{\varphi,\,n=1,M}=j[f_1-f_2]\sin\varphi
\label{S_phi_k_1_M_ser}
\end{equation}
The transpose elements are found by the reciprocity condition $S_{i,j}Z_i=S_{j,i}Z_j$
(see Appendix~\ref{Generalised_S_matrix}, Eq.~\ref{GSM_I_reciprocity}), where $Z_i$ and $Z_j$ are
the impedances for which ports $i$ and $j$ have been defined,
respectively. Note that this reciprocity condition is different from the one used for the voltage calculation in
Section~\ref{Voltage_S_matrix} (see Appendix~\ref{Generalised_S_matrix}).

Given the ports
1,$M$,$\theta$ and $\varphi$ are defined for $Z_0$ of the TL and ports 2 .. $M-1$ are defined for
$Z_T$, the transposed relations $S_{n,\theta}$ and $S_{n,\varphi}$ are given by Eqs.~(\ref{S_theta_k_1_M_ser})
and (\ref{S_phi_k_1_M_ser}), for all $n$:
\begin{equation}
S_{n,\theta}=-j[f_1+f_2]\cos\varphi
\label{S_k_theta_ser}
\end{equation}
\begin{equation}
S_{n,\varphi}=j[f_1-f_2]\sin\varphi
\label{S_k_phi_ser}
\end{equation}
When issuing an excitation $I^+_{\theta}$ from the $\theta$ port and matching all other ports, the currents
at ports $n$ on the TL are $I^-_n=I^+_{\theta}S_{n,\theta}$. Given the number $M$ is not limited, one can get
the continuous current on the TL for the excitation of the $\theta$ port:
\begin{equation}
I_{\theta}(z)=-jI^+_{\theta}[f_1(z)+f_2(z)]\cos\varphi
\label{I_theta}
\end{equation}
and similarly for the $\varphi$ port:
\begin{equation}
I_{\varphi}(z)=jI^+_{\varphi}[f_1(z)-f_2(z)]\sin\varphi
\label{I_phi}
\end{equation}
where the $z$ dependence in $f_{1,2}$ (see Eq.~(\ref{f_12})) is in $l_1$ and $l_2$, defined according
to Eq.~(\ref{l_12_k}), using $z$ for $z_n$.

Expressing those in terms of $V^+_{\theta}=Z_0I^+_{\theta}$, and $V^+_{\varphi}=Z_0I^+_{\varphi}$ results in
Eqs.~(\ref{I_theta_1}) and (\ref{I_phi_1}) in the main text.

The total current developed on the TL is
\begin{equation}
I(z)=I_{\theta}(z)+I_{\varphi}(z)
\label{I_tot_app}
\end{equation}

This concludes the results on a TL matched on both ends. We generalize the above results for TL terminated with
any loads at ports 1 and $M$:
$Z_L$ (left) and $Z_R$ (right), respectively. We express the generalized results in terms of the reflection
coefficients $\Gamma_L$ and $\Gamma_R$, given in Eqs.~(\ref{reflection_coeffs}). For currents they are used with
the minus sign.

At ports 1 and $M$ we have:
\begin{equation}
I^+_1=-\Gamma_L I^-_1\,\,\,\text{and}\,\,\,I^+_M=-\Gamma_R I^-_M,
\label{I_plus_1_M}
\end{equation}
and at ports $n=2..\,M$ (matched with $Z_T$),
\begin{equation}
I^+_{2\le n\le M-1}=0.
\label{I_minus_2_M_minus_1}
\end{equation}
We use the general connection for $1\le n\le M$ (i.e. on TL) but $1\le i\le M+2$,
i.e. all excitations including $\theta$ and $\varphi$:
\begin{align}
I^-_n=&\sum_{i=1}^{M+2} S_{n,i} I^+_i= S_{n,1}I^+_1+S_{n,M}I^+_M+I(z)= \notag \\
       &-S_{n,1}\Gamma_L I^-_1-S_{n,M}\Gamma_R I^-_M+I(z)
\label{I_minus_k}
\end{align}
where the only non zero terms are $i=1,M,\theta$ and $\varphi$ (see Eq~(\ref{I_minus_2_M_minus_1})),
and the last two terms $i=\theta,\varphi$ represent the matched currents at $z=z_n$, given in (\ref{I_theta}),
(\ref{I_phi}), summed in (\ref{I_tot_app}). In the last form of Eq.~(\ref{I_minus_k}), we used (\ref{I_plus_1_M}).

For $n=1$, Eq.~(\ref{I_minus_k}) becomes
\begin{equation}
I^-_1=-\Gamma_R I^-_M e^{-jk2L} + I(-L).
\label{I_minus_1_1}
\end{equation}
because $S_{1,1}=0$, $S_{1,M}=S_{M,1}=e^{-jk2L}$ and $z_n=-L$.

For $n=M$, Eq.~(\ref{I_minus_k}) reads
\begin{equation}
I^-_M=-\Gamma_L I^-_1 e^{-jk2L} + I(L).
\label{I_minus_M_1}
\end{equation}
because $S_{M,M}=0$ and $z_n=L$.

Using $S_{n,1}=e^{-jkl_1}$, $S_{n,M}=e^{-jkl_2}$, for $2\le n\le M-1$, Eq.~(\ref{I_minus_k}) becomes
\begin{equation}
I^-_{2\le n\le M-1}= -e^{-jkl_1}\Gamma_L I^-_1 - e^{-jkl_2} \Gamma_R I^-_M + I(z),
\label{I_minus_k_between_1_and_M_1}
\end{equation}
where in principle $I(z)$ here excludes the terminations, but as we shall see this exclusion is not necessary.

We solve now Eqs.~(\ref{I_minus_1_1}) and (\ref{I_minus_M_1}) for $I^-_1$ and $I^-_M$ and obtain:
\begin{equation}
I^-_1=\frac{-\Gamma_R e^{-j2kL}I(L)+I(-L)}{1-\Gamma_L\Gamma_R e^{-j4kL}} 
\label{I_minus_1_solution}
\end{equation}
\begin{equation}
I^-_M=\frac{-\Gamma_L e^{-j2kL}I(-L)+I(L)}{1-\Gamma_L\Gamma_R e^{-j4kL}}.
\label{I_minus_M_solution}
\end{equation}

We need the total (non matched) current $I_{\text{NM}}(z_n)=I^+_n+I^-_n$ for ports 1 to $M$ on the TL, i.e. the outgoing
$I^-$ plus the incoming $I^+$. For ports 1 or M, this current is
\begin{equation}
I_{\text{NM}}(-L)=I^-_1(1-\Gamma_L)
\label{I_NM_port_1}
\end{equation}
\begin{equation}
I_{\text{NM}}(L)=I^-_M(1-\Gamma_R).
\label{I_NM_port_M}
\end{equation}
For the ports $2\le n\le M-1$, $I^+_n=0$ (\ref{I_minus_2_M_minus_1}),
therefore Eq.~(\ref{I_minus_k_between_1_and_M_1}) describes the total current on those ports. Taking the limit
$z\to -L$ of Eq.~(\ref{I_minus_k_between_1_and_M_1}), using (\ref{I_minus_1_1}) we find it reduces to
(\ref{I_NM_port_1}) and similarly the limit $z\to L$ of Eq.~(\ref{I_minus_k_between_1_and_M_1}) reduces to
(\ref{I_NM_port_M}), therefore Eq.~(\ref{I_minus_k_between_1_and_M_1}) describes
the non matched voltage on the TL at all ports (in the continuum, for all $z$).
Hence, $I_{\text{NM}}(z)= I(z) + \Delta I(z)$, where $\Delta I(z)$ is the correction term due to non matching:
\begin{equation}
\Delta I(z)= -e^{-jkl_1}\Gamma_L I^-_1 - e^{-jkl_2} \Gamma_R I^-_M,
\label{Delta_I_app}
\end{equation}
and $I^-_1$ and $I^-_M$ are given in Eqs.~(\ref{I_minus_1_solution}) and (\ref{I_minus_M_solution}).

\end{document}